%% file: 5ptNMPS.tex
\input harvmacM.tex
\input epsf.sty
\input amssym.tex 

\font\chess=skak10 at 7pt
\def\brook{{\hbox{\chess r}}}
\def\bbishop{{\hbox{\chess b}}}
\def\bknight{{\hbox{\chess n}}}

\def\Tijk#1,#2,#3{T_{#1,#2,#3}}
\def\KN#1{\prod_{j=1}^{#1} e^{k^j\cdot x^j}}
\def\AYM#1{A^{\rm YM}({#1})}
\def\halfap#1{\Big(\mkern-2mu{\ap\over 2}\mkern-2mu\Big)^{\mkern-4mu #1}}
\def\invhalfap#1{\Big(\mkern-1mu{2\over \ap}\mkern-1mu\Big)^{\mkern-4mu #1}}
\def\cI{{\cal I}}
\def\cK{{\cal K}}
\def\cO{{\cal O}}


\preprint DAMTP--2015--20

\title The two-loop superstring five-point amplitude and S-duality

\author Humberto Gomez\email{\bbishop}{humgomzu@ift.unesp.br}, Carlos R. Mafra\email{\brook}{c.r.mafra@damtp.cam.ac.uk} and Oliver Schlotterer\email{\bknight}{olivers@aei.mpg.de}

\address
$^\bbishop$Perimeter Institute for Theoretical Physics, Waterloo, Ontario N2L 2Y5, Canada
$^\bbishop$Instituto de F\'\i sica Te\'orica UNESP --- Universidade Estadual Paulista
Caixa Postal 70532--2 01156--970 S\~ao Paulo, SP, Brazil

\address
$^\brook$DAMTP, University of Cambridge
Wilberforce Road, Cambridge, CB3 0WA, UK

\address
$^\bknight$Max--Planck--Institut f\"ur Gravitationsphysik
Albert--Einstein--Institut, 14476 Potsdam, Germany

\abstract
The low-energy limit of the massless two-loop five-point amplitudes for both type~IIA and
type~IIB superstrings is computed with the pure spinor formalism and its overall coefficient determined
from first principles. For the type~IIB theory, the five-graviton amplitude is found to be proportional
to its tree-level counterpart at the corresponding order in $\ap$. Their ratio ties in with
expectations based on S-duality since it matches the same modular function $E_{5/2}$ which relates the
two-loop and tree-level four-graviton amplitudes. For R-symmetry violating states, the ratio between
tree-level and two-loop amplitudes at the same $\ap$-order carries an additional factor of $-3/5$. Its
S-duality origin can be traced back to a modular form derived from $E_{5/2}$.

\Date {April 2015}


\lref\dhokerReview{
 E.~D'Hoker and D.H.~Phong,
 ``The Geometry of String Perturbation Theory,''
  Rev.\ Mod.\ Phys.\  {\bf 60}, 917 (1988).
}

\lref\GreenYXA{
  M.~B.~Green, S.~D.~Miller and P.~Vanhove,
  ``SL(2,$\Bbb Z$)-invariance and D-instanton contributions to the $D^6 R^4$ interaction,''
[arXiv:1404.2192 [hep-th]].
}

\lref\StiebergerWK{
  S.~Stieberger and T.R.~Taylor,
  ``NonAbelian Born-Infeld action and type 1. - heterotic duality 2: Nonrenormalization theorems,''
Nucl.\ Phys.\ B {\bf 648}, 3 (2003).
[hep-th/0209064].
}
\lref\dhokerVI{
  E.~D'Hoker and D.~H.~Phong,
  ``Two-loop superstrings VI: Non-renormalization theorems and the 4-point function,''
Nucl.\ Phys.\ B {\bf 715}, 3 (2005).
[hep-th/0501197].
}
\lref\PierreFive{
  N.~E.~J.~Bjerrum-Bohr and P.~Vanhove,
  ``Explicit Cancellation of Triangles in One-loop Gravity Amplitudes,''
JHEP {\bf 0804}, 065 (2008).
[arXiv:0802.0868 [hep-th]].
}
\lref\MafraGJA{
  C.~R.~Mafra and O.~Schlotterer,
  ``Towards one-loop SYM amplitudes from the pure spinor BRST cohomology,''
Fortsch.\ Phys.\  {\bf 63}, no. 2, 105 (2015).
[arXiv:1410.0668 [hep-th]].
}
\lref\vafa{
  M.~Bershadsky, S.~Cecotti, H.~Ooguri and C.~Vafa,
  ``Kodaira-Spencer theory of gravity and exact results for quantum string amplitudes,''
Commun.\ Math.\ Phys.\  {\bf 165}, 311 (1994).
[hep-th/9309140].
}
\lref\tsuchiya{
  A.~Tsuchiya,
  ``More on One Loop Massless Amplitudes of Superstring Theories,''
Phys.\ Rev.\ D {\bf 39}, 1626 (1989).
}
\lref\montag{
  J.~L.~Montag,
  ``The one loop five graviton scattering amplitude and its low-energy limit,''
Nucl.\ Phys.\ B {\bf 393}, 337 (1993).
[hep-th/9205097].
}
\lref\RichardsJG{
  D.M.~Richards,
  ``The One-Loop Five-Graviton Amplitude and the Effective Action,''
JHEP {\bf 0810}, 042 (2008).
[arXiv:0807.2421 [hep-th]].
}
\lref\sakai{
  N.~Sakai and Y.~Tanii,
  ``One Loop Amplitudes And Effective Action In Superstring Theories,''
Nucl.\ Phys.\ B {\bf 287}, 457 (1987).
}
\lref\GSoneloop{
  M.B.~Green and J.~H.~Schwarz,
  ``Supersymmetrical Dual String Theory. 3. Loops and Renormalization,''
Nucl.\ Phys.\ B {\bf 198}, 441 (1982).
}
\lref\BjerrumBohrHN{
  N.~E.~J.~Bjerrum-Bohr, P.~H.~Damgaard, T.~Sondergaard and P.~Vanhove,
  ``The Momentum Kernel of Gauge and Gravity Theories,''
JHEP {\bf 1101}, 001 (2011).
[arXiv:1010.3933 [hep-th]].
}

\lref\Enriquez{
  B.~Enriquez,
  ``Analogues elliptiques des nombres multiz\'etas,''
[arXiv:1301.3042 [hep-th]].
}

\lref\BroedelVLA{
  J.~Br\"odel, C.R.~Mafra, N.~Matthes and O.~Schlotterer,
  ``Elliptic multiple zeta values and one-loop superstring amplitudes,''
[arXiv:1412.5535 [hep-th]].
}

\lref\yuri{
  Y.~Aisaka and N.~Berkovits,
  ``Pure Spinor Vertex Operators in Siegel Gauge and Loop Amplitude Regularization,''
JHEP {\bf 0907}, 062 (2009).
[arXiv:0903.3443 [hep-th]].
}

\lref\oneloopbb{
	C.R.~Mafra and O.~Schlotterer,
	``The Structure of n-Point One-Loop Open Superstring Amplitudes,''
	JHEP {\bf 1408}, 099 (2014).
	[arXiv:1203.6215 [hep-th]].
}

\lref\superpoincare{
  N.~Berkovits,
  ``Super-Poincare covariant quantization of the superstring,''
  JHEP {\bf 0004}, 018 (2000)
  [arXiv:hep-th/0001035].
}
\lref\multiloop{
  N.~Berkovits,
  ``Multiloop amplitudes and vanishing theorems using the pure spinor
  formalism for the superstring,''
  JHEP {\bf 0409}, 047 (2004)
  [arXiv:hep-th/0406055].
}

\lref\DHokerEEA{
  E.~D'Hoker and M.~B.~Green,
  ``Zhang-Kawazumi Invariants and Superstring Amplitudes,''
[arXiv:1308.4597 [hep-th]].
\semi
  E.~D'Hoker, M.~B.~Green, B.~Pioline and R.~Russo,
  ``Matching the $D^{6}R^{4}$ interaction at two-loops,''
JHEP {\bf 1501}, 031 (2015).
[arXiv:1405.6226 [hep-th]].
}

\lref\supertwistor{
  N.~Berkovits,
  ``Ten-Dimensional Super-Twistors and Super-Yang-Mills,''
JHEP {\bf 1004}, 067 (2010).
[arXiv:0910.1684 [hep-th]].
}
\lref\NMPS{
  N.~Berkovits,
  ``Pure spinor formalism as an N = 2 topological string,''
  JHEP {\bf 0510}, 089 (2005)
  [arXiv:hep-th/0509120].
}
\lref\humberto{
  H.~Gomez,
  ``One-loop Superstring Amplitude From Integrals on Pure Spinors Space,''
  JHEP {\bf 0912}, 034 (2009)
  [arXiv:0910.3405 [hep-th]].
}
\lref\PSanomaly{
  N.~Berkovits and C.~R.~Mafra,
  ``Some superstring amplitude computations with the non-minimal pure spinor
  formalism,''
  JHEP {\bf 0611}, 079 (2006)
  [arXiv:hep-th/0607187].
}
\lref\FORM{
	J.A.M.~Vermaseren,
	``New features of FORM,''
	[arXiv:math-ph/0010025].
}

\lref\oneloopNMPS{
  C.R.~Mafra and C.~Stahn,
  ``The One-loop Open Superstring Massless Five-point Amplitude with the
  Non-Minimal Pure Spinor Formalism,''
  JHEP {\bf 0903}, 126 (2009)
  [arXiv:0902.1539 [hep-th]].
}
\lref\ictp{
  N.~Berkovits,
  ``ICTP lectures on covariant quantization of the superstring,''
  arXiv:hep-th/0209059.
}

\lref\twoloop{
  N.~Berkovits,
  ``Super-Poincare covariant two-loop superstring amplitudes,''
JHEP {\bf 0601}, 005 (2006).
[hep-th/0503197].
}

\lref\coefftwo{
	H.~Gomez, C.R.~Mafra,
	``The Overall Coefficient of the Two-loop Superstring Amplitude Using Pure Spinors,''
	JHEP {\bf 1005}, 017 (2010).
	[arXiv:1003.0678 [hep-th]].
}
\lref\thetaSYM{
  	J.P.~Harnad and S.~Shnider,
	``Constraints And Field Equations For Ten-Dimensional Superyang-Mills
  	Theory,''
  	Commun.\ Math.\ Phys.\  {\bf 106}, 183 (1986).
\semi
	P.A.~Grassi and L.~Tamassia,
        ``Vertex operators for closed superstrings,''
        JHEP {\bf 0407}, 071 (2004)
        [arXiv:hep-th/0405072].
\semi
	G.~Policastro and D.~Tsimpis,
	``$R^4$, purified,''
	Class.\ Quant.\ Grav.\  {\bf 23}, 4753 (2006).
	[arXiv:hep-th/0603165].
}
\lref\refSYM{
	E.Witten,
        ``Twistor-Like Transform In Ten-Dimensions''
        Nucl.Phys. B {\bf 266}, 245~(1986)
}

\lref\treebbI{
	C.R.~Mafra, O.~Schlotterer and S.~Stieberger,
	``Complete N-Point Superstring Disk Amplitude I. Pure Spinor Computation,'' Nucl.\ Phys.\ B {\bf 873}, 419 (2013).
	[arXiv:1106.2645 [hep-th]].
\semi
	C.R.~Mafra, O.~Schlotterer and S.~Stieberger,
	``Complete N-Point Superstring Disk Amplitude II. Amplitude and Hypergeometric Function Structure,'' Nucl.\ Phys.\ B {\bf 873}, 461 (2013).
	[arXiv:1106.2646 [hep-th]].
}

\lref\PSS{
	C.R.~Mafra,
	``PSS: A FORM Program to Evaluate Pure Spinor Superspace Expressions,''
	[arXiv:1007.4999 [hep-th]].
}
\lref\mafraids{
  C.R.~Mafra,
  ``Pure Spinor Superspace Identities for Massless Four-point Kinematic Factors,''
JHEP {\bf 0804}, 093 (2008).
[arXiv:0801.0580 [hep-th]].
}
\lref\towards{
  C.R.~Mafra,
  ``Towards Field Theory Amplitudes From the Cohomology of Pure Spinor Superspace,''
JHEP {\bf 1011}, 096 (2010).
[arXiv:1007.3639 [hep-th]].
}
\lref\threeloop{
  H.~Gomez and C.R.~Mafra,
  ``The closed-string 3-loop amplitude and S-duality,''
  JHEP {\bf 1310}, 217 (2013).
[arXiv:1308.6567 [hep-th]].
}

\lref\motivic{
  O.~Schlotterer and S.~Stieberger,
  ``Motivic Multiple Zeta Values and Superstring Amplitudes,''
J.\ Phys.\ A {\bf 46}, 475401 (2013).
[arXiv:1205.1516 [hep-th]].
}

\lref\tese{
  C.R.~Mafra,
  ``Superstring Scattering Amplitudes with the Pure Spinor Formalism,''
[arXiv:0902.1552 [hep-th]].
}

\lref\FiveSdual{
  M.B.~Green, C.R.~Mafra and O.~Schlotterer,
  ``Multiparticle one-loop amplitudes and S-duality in closed superstring theory,''
JHEP {\bf 1310}, 188 (2013).
[arXiv:1307.3534 [hep-th]].
}
\lref\EOMBBs{
  C.R.~Mafra and O.~Schlotterer,
  ``Multiparticle SYM equations of motion and pure spinor BRST blocks,''
JHEP {\bf 1407}, 153 (2014).
[arXiv:1404.4986 [hep-th]].
}
\lref\HighSYM{
  C.R.~Mafra and O.~Schlotterer,
  ``A solution to the non-linear equations of D=10 super Yang-Mills theory,''
[arXiv:1501.05562 [hep-th]].
}
\lref\GreenOneLoopcoeff{
M.B.~Green and P.~Vanhove,
  ``The Low-energy expansion of the one loop type II superstring amplitude,''
Phys.\ Rev.\ D {\bf 61}, 104011 (2000).
[hep-th/9910056].
\semi
  M.B.~Green, J.G.~Russo and P.~Vanhove,
  ``Low-energy expansion of the four-particle genus-one amplitude in type II
  superstring theory,''
  JHEP {\bf 0802}, 020 (2008)
  [arXiv:0801.0322 [hep-th]].
\semi
E.~D'Hoker, M.B.~Green and P.~Vanhove,
  ``On the modular structure of the genus-one Type II superstring low-energy expansion,''
[arXiv:1502.06698 [hep-th]].
}

\lref\twolooptwo{
  N.~Berkovits and C.R.~Mafra,
  ``Equivalence of two-loop superstring amplitudes in the pure spinor and RNS formalisms,''
Phys.\ Rev.\ Lett.\  {\bf 96}, 011602 (2006).
[hep-th/0509234].
}

\lref\GGRq{
	M.B.~Green and M.~Gutperle,
	``Effects of D instantons,''
	Nucl.\ Phys.\ B {\bf 498}, 195 (1997).
	[hep-th/9701093].
\semi
	M.B.~Green, M.~Gutperle and P.~Vanhove,
	``One loop in eleven-dimensions,''
	Phys.\ Lett.\ B {\bf 409}, 177 (1997).
	[hep-th/9706175].
}
\lref\GreenKVan{
	M.B.~Green, H.-h.~Kwon and P.~Vanhove,
	``Two loops in eleven-dimensions,''
	Phys.\ Rev.\ D {\bf 61}, 104010 (2000).
	[hep-th/9910055].
}
\lref\Greenthreeloop{
	M.B.~Green and P.~Vanhove,
	``Duality and higher derivative terms in M theory,''
	JHEP {\bf 0601}, 093 (2006).
	[arXiv:hep-th/0510027].
}

\lref\WWWps{
	C.R.~Mafra, O.~Schlotterer,
http://www.damtp.cam.ac.uk/user/crm66/SYM/pss.html
}

\lref\WWWalpha{
	J.~Br\"odel, O.~Schlotterer, S.~Stieberger,
http://mzv.mpp.mpg.de
}

\lref\MafraJQ{
  C.R.~Mafra, O.~Schlotterer, S.~Stieberger and D.~Tsimpis,
  ``A recursive method for SYM n-point tree amplitudes,''
Phys.\ Rev.\ D {\bf 83}, 126012 (2011).
[arXiv:1012.3981 [hep-th]].
}

\lref\onehalfOneloop{
  E.~D'Hoker and D.~H.~Phong,
  ``Multiloop Amplitudes for the Bosonic Polyakov String,''
Nucl.\ Phys.\ B {\bf 269}, 205 (1986).
}

\lref\brenno{
  N.~Berkovits and B.C.~Vallilo,
  ``Consistency of super-Poincare covariant superstring tree amplitudes,''
JHEP {\bf 0007}, 015 (2000).
[hep-th/0004171].
}
\lref\SiegelVol{
	C.L. Siegel, ``Symplectic Geometry'', Am. J. Math. 65 (1943) 1-86;
}
\lref\dhokerS{
  E.~D'Hoker, M.~Gutperle and D.H.~Phong,
  ``Two-loop superstrings and S-duality,''
  Nucl.\ Phys.\  B {\bf 722}, 81 (2005)
  [arXiv:hep-th/0503180].
}
\lref\harris{
Griffiths and Harris, 
``Principles of Algebraic Geometry'', [Wiley Classics Library Edition
Published 1994]
}
\lref\verlinde{
  E.P.~Verlinde and H.L.~Verlinde,
  ``Chiral bosonization, determinants and the string partition function,''
  Nucl.\ Phys.\  B {\bf 288}, 357 (1987).
}

\lref\HullYS{
  C.~M.~Hull and P.~K.~Townsend,
  ``Unity of superstring dualities,''
Nucl.\ Phys.\ B {\bf 438}, 109 (1995).
[hep-th/9410167].
}

\lref\WittenCIA{
  E.~Witten,
  ``More On Superstring Perturbation Theory,''
[arXiv:1304.2832 [hep-th]].
}

\lref\siegel{
	W.~Siegel,
	``Classical Superstring Mechanics,''
	Nucl.\ Phys.\  {\bf B263}, 93 (1986).
}
\lref\teightG{
  M.B.~Green and J.~H.~Schwarz,
  ``Supersymmetrical Dual String Theory. 2. Vertices and Trees,''
  Nucl.\ Phys.\ B {\bf 198}, 252 (1982)..
}
\lref\medina{
L.~A.~Barreiro and R.~Medina,
  ``5-field terms in the open superstring effective action,''
JHEP {\bf 0503}, 055 (2005).
[hep-th/0503182].
}
\lref\KLTref{
  H.~Kawai, D.C.~Lewellen and S.H.H.~Tye,
  ``A Relation Between Tree Amplitudes of Closed and Open Strings,''
Nucl.\ Phys.\ B {\bf 269}, 1 (1986).
}
\lref\Rviolating{
  M.B.~Green, M.~Gutperle and H.~h.~Kwon,
  ``Sixteen fermion and related terms in M theory on $T^2$,''
Phys.\ Lett.\ B {\bf 421}, 149 (1998).
[hep-th/9710151].
\semi
  M.~B.~Green,
  ``Interconnections between type II superstrings, M theory and N=4 supersymmetric Yang-Mills,''
Lect.\ Notes Phys.\  {\bf 525}, 22 (1999).
[hep-th/9903124].
\semi
  N.~Berkovits and C.~Vafa,
  ``Type IIB R**4 H**(4g-4) conjectures,''
Nucl.\ Phys.\ B {\bf 533}, 181 (1998).
[hep-th/9803145].
\semi
A.~Basu and S.~Sethi,
  ``Recursion Relations from Space-time Supersymmetry,''
JHEP {\bf 0809}, 081 (2008).
[arXiv:0808.1250 [hep-th]].
\semi
  R.~H.~Boels,
  ``Maximal R-symmetry violating amplitudes in type IIB superstring theory,''
Phys.\ Rev.\ Lett.\  {\bf 109}, 081602 (2012).
[arXiv:1204.4208 [hep-th]].
\semi
  A.~Basu,
  ``The structure of the $R^8$ term in type IIB string theory,''
Class.\ Quant.\ Grav.\  {\bf 30}, 235028 (2013).
[arXiv:1306.2501 [hep-th]].
}

\lref\stieclosed{
  S.~Stieberger,
  ``Constraints on Tree-Level Higher Order Gravitational Couplings in Superstring Theory,''
Phys.\ Rev.\ Lett.\  {\bf 106}, 111601 (2011).
[arXiv:0910.0180 [hep-th]].
\semi
S.~Stieberger,
  ``Open \& Closed vs. Pure Open String Disk Amplitudes,''
[arXiv:0907.2211 [hep-th]].
}
\lref\stieFive{
  S.~Stieberger and T.~R.~Taylor,
  ``Multi-Gluon Scattering in Open Superstring Theory,''
Phys.\ Rev.\ D {\bf 74}, 126007 (2006).
[hep-th/0609175].
}

\listtoc
\writetoc
\filbreak


\newsec{Introduction}

In this paper, we determine the low-energy limit of the five-point closed-string amplitudes among
massless type IIA and type IIB states using the pure spinor (PS) formalism \refs{\superpoincare,
\NMPS}. The precise elaboration of overall coefficients confirms the predictions
\refs{\GreenKVan,\RichardsJG} based on the non-perturbative S-duality of the type IIB effective action
\HullYS. This complements previous S-duality analyses of the five-point amplitudes at one-loop
\FiveSdual\ as well as the four-point amplitudes at two- \refs{\dhokerS,\coefftwo} and
three-loops \threeloop.

S-duality constrains curvature
couplings of the schematic form $D^{2k} R^n$ (and their supersymmetric completions) to depend on the
scalar fields through modular invariant functions and thereby relates different loop orders in
perturbation theory. The subsequent two-loop analysis probes the moduli-dependent coefficient of the
$D^4R^4$ and $D^2 R^5$ interactions which was identified as the non-holomorphic Eisenstein series
$E_{5/2}$ in ten dimensions \refs{\GreenKVan,\RichardsJG}.
Its perturbative terms relate the tree-level and
two-loop contributions of the corresponding graviton amplitudes and their R-symmetry conserving
superpartners. 

Likewise, R-symmetry violating closed-string amplitudes at different loop-orders 
(involving, for instance, four gravitons and one dilaton) are interlocked by modular forms 
\Rviolating. Given that R-symmetry violating four-point amplitudes vanish, the five-point 
amplitudes in this work furnish the simplest perturbative fingerprints of their modular 
properties. Specifically, the tree-level and two-loop results at the $\ap$-order under 
discussion are expected to orginate from a certain modular derivative of $E_{5/2}$.

We verify the expected ratios by explicit computation at the five-point level, i.e. by
extracting the type IIB components involving five gravitons as well as four gravitons and one dilaton from the supersymmetric
two-loop low-energy limit. This is the first perturbative check at genus two for the S-duality properties of the five-point
interaction $D^2 R^5$ and its R-symmetry violating counterparts.

Since the main objective of this work requires precise control over normalizations, section~2 contains
a detailed account on the conventions used (closely following \refs{\coefftwo, \threeloop}).
In sections~3 and~4, well-known amplitudes at tree-level and one-loop are recomputed using
the conventions of section~2 -- not only to review their end result but
also to verify the reliability of the PS setup in keeping track of their overall normalizations.
The novel result on the two-loop five-point amplitude is derived in section~5.
Finally, section~6 is devoted to the S-duality analysis of the above results and is
suitable for self-contained reading.

\newsec{Review of conventions}
\par\seclab\secNorm

\noindent In this section the conventions used in the rest of the paper are presented. They closely
follow the conventions used in \refs{\coefftwo,\threeloop} but deviations were taken when deemed
appropriate.

\subsec World-sheet fields

The world-sheet action for the left-moving sector in the non-minimal pure spinor formalism is \NMPS\
\eqn\action{
S = {1\over 2\pi \ap}\int d^2z \( \p x^m \pb x_m + \ap p_\a \pb \t^\a
- \ap w_\a \pb\l^\a
- \ap \wb^\a \pb\lb_\a + \ap s^\a\pb r_\a \),
}
where $m=0,1 \ldots,9$ and $\a=1, \ldots,16$ are the vector and spinorial indices of the ten-dimensional
Lorentz group, and $\ap$ denotes the inverse string tension. In addition, $\l^\a$ and $\lb_\a$ are
bosonic pure spinors and $r_\a$ is a constrained fermionic variable,
\eqn\PSconstraints{
(\l\g^m\l) = 0, \quad (\lb\g^m\lb) = 0, \quad (\lb\g^m r) = 0.
}
The Green--Schwarz constraint $d_\a(z)$ and the supersymmetric momentum $\Pi^m(z)$ are defined by
\eqn\dalpha{
d_\a(z) = p_\a - {1\over \ap}(\g^m\t)_\a \p x_m - {1\over 4\ap}(\g^m \t)_\a(\t \g_m \p\t), \quad
\Pi^m(z) = \p x^m + \half (\t\g^m \p\t) \, ,
}
while the BRST charge and the energy-momentum tensor,
\eqn\BRST{
Q=\oint (\l^\a d_\a + \wb^{\a}r_{\a}),\quad
T(z)=-{1\over \ap}\p x^m \p x_m -
p_\a \p\t^\a + w_{\a}\p\l^{\a} +
\wb^\a \p\lb_{\a} -s^\a\p r_{\a} \, ,
}
are related by $\{Q,b(z)\} = T(z)$, with the following expression for the $b$-ghost \NMPS
\eqnn\bghos
$$\eqalignno{
&\qquad b  = s^\a \p\lb_\a +
{1\over 4(\l\lb)}\bigl[ 2\Pi^{m}(\lb \g_{m}d) -
N_{mn}(\lb \g^{mn}\p\t) - J_{\lambda}(\lb \p\t) - (\lb \p^2 \t) \bigr] &\bghos\cr
&{} + {(\lb\g^{mnp}r)\over 192 (\l\lb)^2}\Bigl[ {\ap\over 2}(d\g_{mnp}d) + 24N_{mn}\Pi_{p}\Bigr]
- {\ap\over 2}{(r\g_{mnp}r) \over 16(\l\lb)^3 }\Bigl[
(\lb\g^{m}d)N^{np}
- {(\lb\g^{pqr}r)N^{mn}N_{qr} \over 8(\l\lb)}\Bigr] \, .
}$$

\subsec Scalar Green function and OPEs

The regularized scalar Green function
$G(z,w)$ is written in terms of the prime form $E(z,w)$ and the global holomorphic one-forms  $\omega_I(z)$ as \dhokerReview
\eqn\primeform{
G(z, w) = -{\ap\over2}\ln\big| E(z,w)\big|^2
+\ap\pi\sum_{I,J=1}^g (\Im\!\int_{z}^{w}\!\!\! \omega_I)\,(\Im\Omega)^{-1}_{IJ}\,(\Im\!\int_{z}^{w}\!\!\!\omega_J),
}
and satisfies
\eqnn\Geqs
$$\eqalignno{
{2\over \ap}\p_{z} \pb_{\bar z}G(z,w) &= -2\pi\d^{(2)}(z-w) + \pi  \sum_{I,J=1}^g  \omega_I(z)(\Im\Omega)_{IJ}^{-1}\bar \omega_J(\bar z)&\Geqs\cr
{2\over \ap}\p_{z} \pb_{\bar w}G(z,w) &= 2\pi\d^{(2)}(z-w) - \pi 
 \sum_{I,J=1}^g  \omega_I(z)(\Im\Omega)_{IJ}^{-1}\bar \omega_J(\bar w) \, ,
}$$
where $\Omega_{IJ}$ is the genus-$g$ period matrix to be defined in section~\secMod. Furthermore,
\eqn\defeta{
\eta(z_i,z_j)\equiv \eta_{ij}\equiv -{2\over \ap}{\p\over \p z_i}G(z_i,z_j).
}
The genus-$g$ OPEs are \refs{\verlinde,\siegel}
\eqn\opedp{\eqalign{
x^m(z,\bar z)\,x_n(w,\bar w) &\sim \d^m_n G(z,w), \cr
 d_\a (z)d_\b(w) &\sim - {2\over \ap}\g^m_{\a\b}\Pi_m \eta(z,w),\cr
d_\a(z)\Pi^m(w) &\sim \g^{m}_{\a\b}\p\t^\b \eta(z,w),
}\quad
\eqalign{
p_\a(z)\, \t^\b(w) &\sim \d^\b_\a\eta(z,w),\cr
d_\a(z)f(x(w),\t(w)) &\sim D_{\a}f \eta(z,w),\cr
\Pi^m(z)f(x(w),\t(w)) &\sim -{\ap\over 2}k^m f \eta(z,w) \, ,
}}
where
$D_\a = {\p\over \p\t^\a} + \half (\g^m\t)_\a k_m$
is the supersymmetric derivative and $f(x,\t)$ represents a generic superfield. 

It follows from \opedp\ and \dalpha\ that
\eqn\PiPibar{
\Pi^m(z){\bar\Pi}^n(\bar w) \sim {\ap\over2}\eta^{mn}\Big(
2\pi\d^{(2)}(z-w) - \pi  \sum_{I,J=1}^g  \omega_I(z)(\Im\Omega)_{IJ}^{-1}\bar \omega_J(\bar w)\Big).
}
Left- and right-movers can be kept separated in the evaluation of the amplitude
by expanding
$\Pi^m(z) = \hat\Pi^m(z) + \sum_{I=1}^g \Pi^m_I \omega_I(z)$ and computing the holomorphic square with
\eqn\LRcontract{
\Pi^m_I \bar\Pi^n_J = -{\ap\over2} \eta^{mn}\pi\,(\Im\Omega)_{IJ}^{-1}.
}
Using this prescription, contributions containing a single $\Pi^m_I$ or
$\bar\Pi^m_I$ vanish.

\subsec SYM superfields and massless vertex operators

The closed-string massless vertex operators are related to the holomorphic square of the open string
vertex operators
\eqn\vertices{
V = \l^\a A_\a(x,\t), \quad U = \p\t^\a A_\a(x,\t) + \Pi^m  A_m(x,\t) + {\ap\over 2}d_\a W^\a(x,\t)
+ {\ap \over 4}N_{mn} F^{mn}(x,\t) \, ,
}
where
$A_\a(x,\t),A^m(x,\t),W^\a(x,\t)$ and $F^{mn}(x,\t)$ are the super-Yang--Mills (SYM) superfields in
ten dimensions. Their equations of motion \refSYM
\eqnn\SYM
$$\displaylines{
\hfill D_\a A_\b + D_\b A_\a = \g^m_{\a\b} A_m, \qquad D_\a A_m = (\g_m W)_\a + k_m A_\a  \hfill\phantom{(1.1)}\cr
\hfill D_\a F_{mn} = 2k_{[m} (\g_{n]} W)_\a, \qquad  D_\a W^{\b} = {1\over 4}(\g^{mn})_\a{}^\b F_{mn}  \hfill\SYM\cr
}$$
are solved by the $\theta$-expansions in \thetaSYM\ involving gluon polarization vectors and gaugino wave functions.
More precisely, the closed-string vertex operators are given by
\eqn\holpres{
|V(z)|^2 \equiv  V(\t)\otimes \tilde V(\bar\t) e^{k\cdot x}, \quad
|U(z)|^2 \equiv  U(\t)\otimes \tilde U(\bar\t) e^{k\cdot x} \, ,
}
where $V(\t)$ and $U(\t)$ are defined from \vertices\ by stripping off the plane-wave factor, e.g.
$U(z) = U(\t)e^{k\cdot x}$. Furthermore, each massless vertex is normalized with a coefficient $\k$
(see e.g. \dhokerS) so the $n$-point
amplitude prescription contains an overall factor of $\k^n$. As shown in appendix~\appUNI, unitarity
relates it to the other string parameters (such as the coupling constant $e^{-2\lambda}$) via $\k^2
e^{-2\l} = \pi/\ap^2$.

\subsec Integration on pure spinor space

The zero-mode measures for the non-minimal pure spinor variables in a genus-$g$ surface
have length dimension zero and
are given by \coefftwo
\eqn\measures{\eqalign{
&[d\l]\, T_{\a_1 \ldots\a_5} = c_{\l}\, \e_{\a_1 \ldots\a_{16}} d\l^{\a_6}\kern-4pt \ldots d\l^{\a_{16}} \cr
&[d\lb]\, {\bar T}^{\a_1 \ldots\a_5} \kern-2pt = c_{\lb}\, \e^{\a_1 \ldots\a_{16}} d\lb_{\a_6} \ldots d\lb_{\a_{16}} \cr
&[dr] = c_r\, {\bar T}^{\a_1 \ldots\a_5} \e_{\a_1 \ldots\a_{16} } \p_{r}^{\a_6} \ldots \p_{r}^{\a_{16}}\cr
&[d\t] = c_\t\, d^{16}\t
}\qquad\eqalign{
&[dw] = c_{w}\, T_{\a_1 \ldots\a_5} \e^{\a_1 \ldots\a_{16}} dw_{\a_6} \ldots dw_{\a_{16}}\cr
&[d\wb]\, T_{\a_1 \ldots\a_5} = c_{\wb}\, \e_{\a_1 \ldots\a_{16}} d\wb^{\a_6}\! \ldots d\wb^{\a_{16}}\cr
&[ds^I] = c_s\, T_{\a_1 \ldots\a_5} \epsilon^{\a_1\ldots \a_{16}} \p^{s^I}_{\a_6}\ldots \p^{s^I}_{\a_{16}}\cr
&[dd^I] = c_d\, d^{16}d^I.
}}
The normalizations are \coefftwo
\eqn\normalizations{\eqalign{
c_{\l} &= \halfap{-2}\mkern-4mu {1 \over 11!} \Big({A_g \over 4\pi^2}\Big)^{\!\! 11/2}\cr
c_{\lb} &= \halfap{2} {2^6 \over 11!} \Big({A_g \over 4\pi^2}\Big)^{\mkern-6mu 11/2}\cr
c_r &= \halfap{-2}\mkern-10mu {R \over 11!5!}\Big({2\pi \over A_g}\Big)^{\mkern-6mu 11/2}\cr
c_{\t} &= \halfap{4}\mkern-4mu\Big({2\pi\over A_g}\Big)^{\mkern-6mu 16/2}\cr
}\qquad\eqalign{
c_{w} &= \halfap{2} {(2\pi)^{-11}\over 11!\,5!}\, Z_g^{-11/g}\cr
c_{\wb} &= \halfap{-2} \mkern-16mu {(\l\lb)^3\over 11!\,(2\pi)^{11}} Z_g^{-11/g}\cr
c_s &= \halfap{2} {(2\pi)^{11/2} R^{-1} \over 2^6 11!\,5!\, (\l\lb)^3} Z_g^{11/g}\cr
c_{d} &= \halfap{-4} \!\! (2\pi)^{16/2}\, Z_g^{16/g}\,, \cr
}}
where $A_g = \int d^2z \sqrt{h}$ denotes the area of the genus-$g$ Riemann surface with metric $h$,
\eqn\Zg{
 \quad Z_g = {1\over \sqrt{\det(2\ImOmega)}}, \qquad g \ge 1\,,
}
and $R$ is an arbitrary parameter capturing the freedom to normalize the
string tree-level amplitudes\foot{In previous works \refs{\coefftwo,\threeloop} the
choice $R=\sqrt{2}/(2^{16}\pi)$ was made to match the tree-level conventions of \dhokerS.
In this work we deviate from that motivation and the choice \convention\ will lead to
tree-level amplitudes \treeFin\ with unit overall coefficient.}.
As discussed in \coefftwo, the final expressions for multiloop amplitudes are independent of the area
$A_g$. The tensors $T_{\a_1\ldots \a_5}$ and ${\bar T}^{\a_1\ldots \a_5}$ appearing in \measures\
are totally antisymmetric due to the pure spinor constraint \PSconstraints,
\eqnn\Ttensors
$$\eqalignno{
T_{\a_1\a_2\a_3\a_4\a_5} &=
(\l \g^m)_{\a_1}(\l \g^n)_{\a_2}(\l \g^p)_{\a_3} (\g_{mnp})_{\a_4\a_5}\,, &\Ttensors\cr
{\bar T}^{\a_1\a_2\a_3\a_4\a_5} &= (\lb \g^m)^{\a_1}(\lb \g^n)^{\a_2}(\lb \g^p)^{\a_3} (\g_{mnp})^{\a_4\a_5}
}$$
and satisfy $T\cdot {\bar T} = 5!\, 2^6 (\l\lb)^3$.

One can show using the results of \humberto\ that the integration over
an arbitrary number of pure spinors $\l^\a$ and $\lb_\b$ is given by
\eqn\dladlb{
\int[d\l][d\lb]e^{-(\l\lb)}(\l\lb)^m \l^{\a_1}\cdots \l^{\a_n}\lb_{\b_1}\cdots \lb_{\b_n} =
\Big({A_g\over 2\pi}\Big)^{\mkern-6mu 11}{\Gamma(8+m+n)\over 302400}{\cal T}{}^{\a_1 \ldots\a_n}_{\b_1 \ldots\b_n} \, ,
}
where ${\cal T}{}^{\a_1 \ldots\a_n}_{\b_1 \ldots\b_n}$ are the $\g$-matrix traceless tensors discussed
in \threeloop.
From ${\cal T}{}^{\a_1 \ldots\a_p}_{\a_1 \ldots\a_p} = 1$ it follows that \coefftwo
\eqn\humbps{
\int [d\l][d\lb] (\l\lb)^n e^{-(\l\lb)} = \Big({A_g\over 2\pi}\Big)^{\mkern-6mu 11}{\Gamma(8+n)\over 7!\, 60}\,.
}
For an arbitrary superfield $M(\l,\lb,\t,r)$ we define \coefftwo
\eqn\save{
\langle M(\l,\lb,\t,r)\rangle_{(p,g)} \equiv
\int [d\t][dr][d\l][d\lb]\, {e^{-(\l\lb)-(r\t)}\over (\l\lb)^{3-p}}\, M(\l,\lb,\t,r)\,,
}
and therefore the
pure spinor measure $(\l\g^r \t)(\l\g^s \t)(\l\g^t \t)(\t\g_{rst}\t)  \equiv  (\l^3\t^5)$
is mapped to
\eqn\Ndef{
\langle (\l^3\t^5) \rangle_{(p,g)}
= N_{(p,g)} \langle(\l^3\t^5)\rangle\,,\quad
N_{(p,g)} \equiv 2^7 {R\over P}\,\Big({ 2\pi\over A_g}\Big)^{\mkern-6mu 5/2}\! \halfap{2}\, {\Gamma(8+p)\over 7!} \, ,
}
and the identity factor ${\langle (\l^3\t^5)\rangle\over P} = 1$ keeps track
of the normalization convention \superpoincare
\eqn\PSSnorm{
\langle (\l\g^r \t)(\l\g^s \t)(\l\g^t \t)(\t\g_{rst}\t)\rangle = P.
}
The choice $P=2880$ is convenient in view of the factorization properties
of pure spinor superspace kinematic factors and has been observed in \brenno\ to
imply tree-level normalizations compatible with RNS computations.
Unless otherwise noted we use,
\eqn\convention{
R^2={\pi^5\over 2^5},\quad P= 2880 \ .
}

\subsubsec Abbreviations and (anti-)symmetrization combinatorics

The (anti)symmetrization over $n$ indices includes a factor of $1/n!$, the generalized
Kronecker delta is $\d^{\a_1 \ldots\a_n}_{\b_1 \ldots \b_n} \equiv \d^{[\a_1}_{\b_1} \cdots \d^{\a_n]}_{\b_n}$ and
satisfies $\d^{\a_1 \ldots\a_n}_{\a_1 \ldots \a_n} = {d\choose n}$ where $d=10$ ($d=16$) for
vector (spinor) indices.
The integration over $\t$ is given by $\int d^{16}\t\, \t^{\a_1} \cdots \t^{\a_{16}} = \e^{\a_1 \ldots \a_{16}}$ and
$\e^{\a_1 \ldots \a_{11}\g_1 \ldots\g_5 }\e_{\a_1 \ldots \a_{11}\b_1 \ldots\b_5} = 11!5!\,\d^{\g_1
\ldots\g_5}_{\b_1 \ldots\b_5}$.

Partitions of $d_\a$ zero-modes are denoted by $(p_1,p_2, \ldots, p_g)_d$, signaling the presence of
$p_I$ factors of $d^I_\a$ for $I=1,2, \ldots, g$. Accordingly, contributions from the $b$-ghost will be
labeled by their partition of
$d_\a$ zero-modes as $B^{m_1 \ldots m_r}_{(p_1,p_2, \ldots,p_g)}$, where the vector indices take into account
that those contributions need not be Lorentz scalars.
Furthermore, we define
\eqnn\defs
$$\eqalignno{
(\e\cdot T\cdot d^I) &\equiv \e^{\a_1 \ldots\a_{16}}T_{\a_1 \ldots\a_5}d^I_{\a_6}\cdots d^I_{\a_{16}}\,,\quad
(\lb r d^Id^J)\equiv  (\lb\g^{mnp}r)(d^I\g_{mnp}d^J)&\defs\cr
}$$
\eqn\definitionD{
D_{(11+p_1,11+p_2,\ldots,11+p_g)}^{m_1m_2\ldots m_r} \equiv
\int \prod_{I=1}^g [dd^I]{(\e\cdot T\cdot d^I)\over 11!\,5!} B_{(p_1,p_2,\ldots,p_g)}^{m_1\ldots m_r} \ .
}

\subsubsec Frequent zero-mode integrals

Some integrals which are frequently used in the next sections are summarized here,
\eqnn\Tds
$$\eqalignno{
\int [dd^I](\e\cdot T\cdot d^I)\,d^I_{\a_1}d^I_{\a_2}d^I_{\a_3}d^I_{\a_4}d^I_{\a_5} &=
11!\,5!\,c_d\,T_{\a_1\a_2\a_3\a_4\a_5} &\Tds\cr
\int [dd^I](\e\cdot T\cdot d^I)\,d^I_{\a_1}d^I_{\a_2}d^I_{\a_3}(d^I\g^{mnp}d^I) &=
11!\,5!\,96\,c_d\,(\l\g^{[m})_{\a_1}(\l\g^n)_{\a_2}(\l\g^{p]})_{\a_3}\,,\cr
\Big|\int \prod_{I=1}^g [dw^I][d\wb^I][ds^I]\, e^{-(w^I\wb^I) - (d^I s^I)}\Big|^2 &=
\halfap{4g}\!\!\!\!{1\over (2\pi)^{16g} 2^{2g} Z_g^{22}}\Big|\prod_{I=1}^g {(\e\cdot T\cdot d^I)\over
(11!\,5!)}\Big|^2 \ .
}$$
To prove the third integral one uses \refs{\humberto,\coefftwo}
\eqnn\dablios
$$\eqalignno{
\int \prod_{I=1}^g [ds^I]\, e^{-(d^Is^I)} &= \halfap{2g}{(2\pi)^{11g/2}Z_g^{11}\over R^g
2^{6g}(\l\lb)^{3g}}
\prod_{I=1}^g {(\e\cdot T\cdot d^I)\over (11!\,5!)}\,,\cr
\int \prod_{I=1}^g [dw^I][d\wb^I]\, e^{-(w^I\wb^I)} &= {(\l\lb)^{3g}\over (2\pi)^{11g}}Z_g^{-22}\,. &\dablios\cr
}
$$

\subsec Riemann surfaces and moduli space
\par\subseclab\secMod

\noindent A holomorphic field with conformal weight one in a genus-$g$ Riemann surface $\Sigma$ can be expanded
in a basis of holomorphic one-forms as
$\phi(z) = {\hat \phi}(z) + \sum_{I=1}^g \omega_I(z)\phi^I$,
and $\phi^I$ are the {\it zero-modes} of $\phi(z)$. If $\{a_I,b_J\}$ are the generators of the $H_1(\Sigma_{g},\Bbb Z)=\Bbb Z^{2g}$
homology group, the holomorphic one-forms can be chosen such that for $I,J=1,2,\ldots,g$
\eqn\RieOmega{
\int_{a_I} \!\omega_J(z)\,dz = \d_{IJ}, \quad
\int_{b_I}\! \omega_J(z)\,dz =\Omega_{IJ},\quad
\int d^2z\, \omega_I\,\bar \omega_J  = 2 \ImOmega_{IJ} \, ,\quad
}
where $\Omega_{IJ}$ is the symmetric period matrix with $g(g+1)/2$ complex degrees
of freedom and $d^2z =i dz\wedge d{\bar z} = 2\,d{\rm Re}(z)d{\rm Im}(z)$ \dhokerReview.
We also define
\eqn\defdel{
\int_{\Sigma_n} \equiv \int \prod_{i=1}^n d^2z_i\,,\quad
\Delta_{ij}\equiv \e^{IJ}\omega_I(z_i)\omega_J(z_j) = \omega_1(z_i) \omega_2(z_j) - \om_1(z_j) \om_2(z_i)\,.
}
The moduli space ${\cal M}_g$ is defined as the space of inequivalent complex structures $\tau_i$
on the Riemann surface of genus $g$ and has complex dimension
$3g-3$, for $g>1$.
For genus two and three, the dimension of the moduli space is the same as the dimension
of the period matrices ($3g-3 = g(g+1)/2$ for $g=2,3$) and the amplitudes can be parameterized by the period
matrix instead of the moduli coordinates; more explicitly for genus two \dhokerReview,
\eqn\mparam{
\int d^2y\, \omega_I(y)\,\omega_J(y) \mu_i(y) = {\delta \Omega_{IJ}\over \delta\tau_i},\quad
\int_{{\cal M}_2} \!\!\!d^2\tau\; \Bigl|\e_{i_1 i_2 i_3}{\delta\Omega_{11}\over \delta\tau_{i_1}}
{\delta\Omega_{12}\over \delta\tau_{i_2}}
{\delta\Omega_{22}\over \delta\tau_{i_3}}\Bigr|^2
=\int_{{\cal F}_2} d^2\Omega \, ,
}
where $d^2\tau \equiv \prod_{j=1}^{3g-3} d^2\tau_j $,
$d^2\Omega \equiv \prod_{I\leq J}^{g} d^2\Omega_{IJ}$ and ${\cal F}_g$ denotes the
fundamental domain of $Sp(2g,{\Bbb Z})/{\Bbb Z}_2$. To avoid cluttering, the domains ${\cal M}_g$ and ${\cal F}_g$
will be henceforth omitted.

The $Sp(2g,{\Bbb Z})$-invariant measure for the genus-$g$ moduli space and its
volume are \SiegelVol\foot{The definition of $d^2\Omega$ here
is $2^{g(g+1)/2}$ bigger than in the original
formula of \SiegelVol.}
\eqn\Volumes{
d\mu_g \equiv {d^2\Omega\over (\det\ImOmega)^{g+1}},\quad
\int d\mu_g = 2 \prod_{k=1}^g \({2^k\over \pi^k}\Gamma(k)\zeta_{2k}\)\,.
}
In particular,
\eqn\AllVolumes{
\int d\mu_1 = {2\pi\over3}, \quad \int d\mu_2={4\pi^3\over 3^3\,5}, \quad \int d\mu_3 = {2^6\pi^6\over 3^6\,5^2\,7}\,.
}

\subsec{The amplitude prescription}

The multiloop $n$-point closed-string amplitude prescription was given in \NMPS\
\eqnn\amplitude
$$\eqalignno{
M_n^{(1)} &=
S_1 \k^n \int  d^2\tau \int_{\Sigma_{n-1}}
\left|\langle \langle {\cal N}^{(1)}  (b,\mu)\, V^1(0)U^2(z_2) \cdots U^n(z_n) \rangle \rangle\right|^2\,, &\amplitude\cr
M_n^{(g)} &=
S_g \k^n e^{(2g -2)\l} \int  \prod_{j=1}^{3g-3} d^2\tau_j \int_{\Sigma_n}
\left|\langle \langle {\cal N}^{(g)}  (b,\mu_j)\, U^1(z_1) \cdots U^n(z_n) \rangle \rangle\right|^2\,,\quad g\ge 2 \ .
}$$
The symmetry factors for the one- and two-loop amplitudes are $S_1=1/2$ \refs{\onehalfOneloop,\sakai}
and $S_2=1/2$ \vafa.
Furthermore, $S_g=1$ for\foot{We thank Edward Witten for emphasizing this point to us.} $g>2$.
The $b$-ghost insertion is
\eqn\binsert{
(b,\mu_j) = {1\over 2\pi}\int d^2 y\, b(y)\mu_j(y), \quad j=1,\ldots ,3g-3 \, ,
}
where $\mu_j$ denotes the Beltrami differential for the modulus parameter $\tau_j$,
and ${\cal N}^{(g)}$ is the BRST regulator \NMPS
\eqn\calNg{
{\cal N}^{(g)} \equiv \exp{\Big(-\llb - (r\t) + \sum_{I=1}^g \big[(w^I\wb^I) + (s^I d^I) \big]\Big)}\,.
}
The bracket $\langle\langle \ldots \rangle\rangle$ in \amplitude\ denotes the path integral which integrates out
the non-zero modes through OPEs and additionally contains the zero-mode integration
measure
\eqn\zeromes{
\langle \ldots \rangle = \int [d\theta][dr][d\l][d\lb] \prod_{I=1}^g [dd^I][ds^I][d\wb^I][dw^I] \ldots
}
After the integration over $[dd^I][ds^I][dw^I][d\wb^I]$ has been performed, the remaining variables
$\l^\a,\lb_\b,\t^\d$ and $r_\a$ have conformal weight zero
and therefore are the same ones which need
to be integrated in the prescription of the tree-level amplitudes. Using the Theorem~1 from \threeloop\
all correlators at this stage of the computation reduce to pure spinor superspace expressions
whose component expansions can be straightforwardly computed\foot{Note that
$r_\a$ variables are converted to $D_\a$ derivatives using
$r_\a e^{-(r \t)} = D_\a e^{-(r \t)}$ \PSanomaly.} \refs{\PSS,\FORM} from the $\theta$-expansions in \thetaSYM. In
particular, the last correlator to evaluate is a combination of the zero-mode integration of tree-level
pure spinor variables \Ndef\ and $x^m$ \coefftwo
\eqn\NKN{
N_{(p,g)}^2 \bigl\langle \prod_{j=1}^{n}e^{ k^j\cdot x^j}\bigr\rangle =
(2\pi)^{10}\d^{10}(k)\halfap{-1} {2^{9}R^2\over \pi^5P^2}\({\Gamma(8+p)\over 7!}\)^2
\cI^{(g)}_n \, ,
}
where $\d^{10}(k) \equiv \d^{10}(\sum_i k^m_i)$ and ${\cal I}_n^{(g)}$ is the $n$-particle
Koba--Nielsen factor
\eqn\defKN{
\cI^{(g)}_n \equiv \exp \Big( \sum_{i<j}^n s_{ij} G_{ij} \Big),\qquad s_{ij}\equiv k_i \cdot k_j\, .
}
Some products which appear in later sections are
\eqnn\Nthreezero
$$\eqalignno{
N_{(3,g)}^2 \bigl\langle \prod_{j=1}^{n}e^{ k^j\cdot x^j}\bigr\rangle &=
(2\pi)^{10}\d^{10}(k)\halfap{-1}\;\cI^{(g)}_n &\Nthreezero\cr
N_{(2,g)}^2 \bigl\langle \prod_{j=1}^{n}e^{ k^j\cdot x^j}\bigr\rangle &=
(2\pi)^{10}\d^{10}(k)\halfap{-1}{1\over 2^2 5^2}\;\cI^{(g)}_n\cr
N_{(0,g)}^2 \bigl\langle \prod_{j=1}^{n}e^{ k^j\cdot x^j}\bigr\rangle &=
(2\pi)^{10}\d^{10}(k)\halfap{-1}{1\over 2^8 3^4 5^2}\;\cI^{(g)}_n\,.\cr
%
}$$
Given the above conventions in \dim, the length dimension $[\ldots]$ of the closed-string $n$-point amplitude is
independent of the genus; $[M_n^{(g)}] = n(2 + [\kappa])$. Since $[\k] = -2$ (see appendix~\appUNI)
the amplitudes are dimensionless.
Furthermore, in most of the calculations below overall minus signs
will not be rigorously tracked.

For four-point amplitudes it is convenient to use the following shorthand notation
for symmetric polynomials in Mandelstam invariants \defKN
\eqn\defsig{
\sigma_k \equiv \halfap{k}( s_{12}^k+s_{13}^k+s_{14}^k)\,.
}

\subsec Multiparticle fields
\par\subseclab\secMP

\noindent The five-point amplitudes at genus $g=1,2$ discussed in this work reconcile zero-mode saturation
with one OPE among the vertex
operators of both left- and right-movers. The systematics of
OPEs has been studied using multiparticle fields in \EOMBBs, starting with:
\eqnn\VUOPE
\eqnn\UUOPE
$$\eqalignno{
V_1(z_1) U_2(z_2) &={|z_{12}|^{-{\alpha' \over 2} s_{12} }\over z_{21} } \halfap{} \big[ V_{12} + Q(\ldots) \big] &\VUOPE \cr
U_1(z_1) U_2(z_2) &={|z_{12}|^{-{\ap\over 2} s_{12} }\over z_{21}} \halfap{} \big[ \partial \theta^\alpha
A^{12}_\alpha+ \Pi_m A^m_{12} + \halfap{} d_\alpha W^{\alpha}_{12} + {\ap\over 4} N_{mn} F^{mn}_{12} \big]\cr
&\quad + \partial_{1 ,2}(\ldots) \ .&\UUOPE
}$$
The suppressed BRST-exact terms in \VUOPE\ and worldsheet derivatives in \UUOPE\ drop out from the
subsequent computations. The two-particle superfields of interest in this work are
\eqnn\twopart
$$\eqalignno{
A_{12}^\alpha&\equiv - \half\bigl[ A^1_\a (k^1\cdot A^2) + A^1_m (\g^m W^2)_\a - (1\leftrightarrow 2)\bigr] \cr
W_{12}^\alpha &\equiv {1\over 4}(\g^{mn}W^2)^\a F^1_{mn} + W_2^\a (k^2\cdot A^1) - (1\leftrightarrow 2)
&\twopart \cr
F_{12}^{mn} &\equiv F_2^{mn}(k^2\cdot A^1) +  F_2^{[m}{}_{p}F^{n]p}_1 + k_{12}^{[m}(W_1\g^{n]}W_2) -
(1\leftrightarrow 2)
}$$
with a similar definition for $A_{12}^m$, and
\eqn\Vtwo{
V_{12} \equiv \lambda^\alpha A^{12}_\alpha \ , \ \ \ \ \ \ Q V_{12} = s_{12} V_{1}V_{2} \ .
}
Generalizations to $p\geq 3$ particles, in particular the $V_{12\ldots p}$ mentioned in the context of
tree amplitudes, can be found in \EOMBBs. In a notation where $A,B,C,\ldots$ denote multiparticle
labels such as $A=12\ldots p$, the simplest class of one-loop kinematic factors are given by
\eqn\Tijkdef{
T_{A,B,C} \equiv {1\over 3}\big[ (\l\g_m W_A) (\l\g_n W_B)F_C^{mn} + (C \leftrightarrow A,B)\big]\, .
}
They were firstly studied in the context of multiparticle open-string amplitudes at one-loop
\oneloopbb\ and identified as box-numerators in one-loop amplitudes of ten-dimensional SYM \MafraGJA.

\subsec Length dimensions

For convenience, the length dimensions of various fields and constants used throughout this work
are summarized here,
\eqnn\dim
$$\displaylines{
[\ap] = 2,\ \; [x^m] = 1, \ \; [k^m] =-1, \ \; [\k] = -2,\ \; [G(z,w)] = 2,\ [\eta_{ij}] = 0 \hfil\dim\hfilneg\cr
[\t^\a, \l^\a, \wb^\a, s^\a] = \half, \quad [p_\a, w_\a, \lb_\a, r_\a] = -\half,\quad [Q]=[b]=[T] = 0, \cr
[A^{12 \ldots p}_\a] = {3\over 2} - p,\ \; [A^{12 \ldots p}_m] = 1 - p,\ \; [W^\a_{12 \ldots p}] =
{1\over 2} - p,\
\; [F^{12 \ldots p}_{mn}] = -p, \cr
[V(z)] = [U(z)] = 1,\quad
[\AYM{1,2, \ldots,n}] = n-4,\quad [\d^{10}(k)] = 10,\quad [M_n^{(g)}] = 0 \ .\cr
%
}$$

\newsec{Tree-level closed-string amplitudes}

In this section the tree-level amplitudes involving $n=3,4,5$ closed-string states
are reviewed and recomputed using the normalization conventions of section~\secNorm. This
ensures that the S-duality discussion of section~\secSdual\ uses amplitudes computed
with a uniform set of conventions (which differ from \refs{\coefftwo,\threeloop}).
For earlier references, see \refs{\teightG,\KLTref,\medina,\stieclosed}.

\subsec The amplitude prescription

The prescription to compute the $n$-point tree-level amplitude in the PS formalism is \NMPS,
\eqn\presc{
M^{(0)}_n = \kappa^n e^{-2\l} \int \prod_{i=2}^{n-2} d^2 z_i |\langle\langle
{\cal N}^{(0)}V_1(0) U_2(z_2) \ldots U_{n-2}(z_{n-2}) V_{n-1}(1)V_n(\infty)\rangle\rangle|^2 \, ,
}
where ${\cal N}^{(0)} = e^{-\llb -r\t}$ is the zero-mode regulator at genus zero. As explained below \calNg,
$\langle\langle \ldots\rangle\rangle$ denotes the path integral which reduces to
the integration over the zero-modes of tree-level variables
after the non-zero modes are integrated out through OPEs. The pure spinor 
computation of the $n$-point tree-level correlator can be found in \treebbI,
\eqnn\MSScorr
$$\eqalignno{
\langle {\cal K}^{(0)}(z_2, \ldots,z_{n-2}) \rangle &\equiv \langle \langle V_1(z_1) U_2(z_2) \cdots U_{n-2}(z_{n-2})
V_{n-1}(z_{n-1})V_n(z_n) \rangle \rangle &\MSScorr\cr
&=\halfap{n-3} \sum_{p=1}^{n-2} { \langle  V_{12 \ldots p}\; V_{n-1,n-2, \ldots, p+1} V_n \rangle
\over (z_{12}z_{23} \cdots z_{p-1,p})(z_{n-1,n-2}\cdots z_{p+2,p+1})} + {\cal P}(2,\ldots,n-2)  \, , \cr
}$$
where ${\cal P}(2, \ldots,n-2)$ instructs to sum over all permutations of $2, \ldots, n-2$. The M\"obius symmetry of the genus-zero
 worldsheet has been fixed by setting $\{z_1,z_{n-1},z_n\} =\{0,1,\infty\}$. The
correlator \MSScorr\ was
later identified as a superposition of SYM tree amplitudes \treebbI\ (see \MafraJQ\ for their
pure spinor superspace representation),
\eqnn\MSSYM
$$\eqalignno{
\langle {\cal K}^{(0)}(z_2, \ldots,z_{n-2}) \rangle &= \halfap{n-3}{s_{12}\over z_{12}} \left(  {s_{13}\over z_{13}}
+ {s_{23}\over z_{23}} \right) \cdots \left( {s_{12}\over z_{12}}+\cdots
+   {s_{1,n-2}\over z_{1,n-2}} \right) &\MSSYM\cr
&\qquad \times A^{\rm YM}(1,2,\ldots,n-1,n)
 + {\cal P}(2,\ldots,n-2)  \ .
}$$
The multiparticle superfields $V_{12}$ and $V_{12\ldots p}$ in \MSScorr\ are defined in
 \Vtwo\ and \EOMBBs, respectively. Therefore, the prescription \presc\ yields,
\eqnn\treeFin
$$\eqalignno{
M^{(0)}_n &= \kappa^n e^{-2\l} \int \prod_{i=2}^{n-2} d^2 z_i
|\langle {\cal K}^{(0)}(z_2, \ldots,z_{n-2})\rangle_{(3,0)}|^2\;  \Big \langle\prod_{j=1}^{n} e^{k^j\cdot x^j} \Big \rangle &\treeFin\cr
 &= (2\pi)^{10}\d^{10}(k)\halfap{-1}\kappa^n e^{-2\l}
 \int \prod_{i=2}^{n-2} d^2 z_i |\langle {\cal K}^{(0)}(z_2, \ldots,z_{n-2})\rangle|^2 \cI^{(0)}_n \ ,
 }$$
where we used \Ndef\ and \Nthreezero. Note that the Koba-Nielsen
factor \defKN\ simplifies to ${\cal I}^{(0)}_n= \prod_{i<j}^n |z_{ij}|^{-\ap s_{ij}}$ at genus zero.

\subsec The three-point amplitude

Using the formula \treeFin\ and taking  $\cI^{(0)}_3=1$ into account, the three-point
amplitude can be written down immediately
\eqn\threeptAmp{
M^{(0)}_3 = (2\pi)^{10}\d^{10}(k) \halfap{-1}\!\kappa^3 e^{-2\l}\cK_3^{(0)} \, ,
}
where $\cK_3^{(0)}\equiv |\langle V_1V_2V_3\rangle|^2 = |\AYM{1,2,3}|^2$ and (note $[\cK_3^{(0)}]=-2$)
\eqn\AYMthree{
\langle V_1 V_2 V_3\rangle = (e^1\cdot e^2)(k^2\cdot e^3) + e^1_m (\chi_2 \gamma^m \chi_3) + {\rm cyc}(1,2,3)\,.
}
The component expressions are derived from the $\theta$-expansions of \thetaSYM\ and involve transverse polarization vectors $e^i$ of the
gluon as well as chiral spinor wave functions $\chi_i$ of the gluino.

\subsec The four-point amplitude

Similarly, using the formula \treeFin\ the four-point amplitude becomes
\eqn\fourtmp{
M^{(0)}_4 = (2\pi)^{10}\d^{10}(k)\halfap{-1}\! \kappa^4 e^{-2\l} \int d^2 z_2 |\langle
{\cal K}^{(0)}(z_2)\rangle|^2 \cI^{(0)}_4 \ ,
}
where the correlator is \treebbI\ (see also \mafraids)
\eqnn\fourMSS
$$\eqalignno{
\langle {\cal K}^{(0)}(z_2)\rangle &= \langle \langle V_1(z_1) U_2(z_2) V_3(z_3) V_4(z_4)\rangle \rangle =
\halfap{}\Bigl[{\langle V_{12}V_3V_4\rangle \over z_{12}} + {\langle V_{1}V_{32}V_4\rangle \over
z_{32}}\Big]&\fourMSS\cr
&= \halfap{}\,{s_{12}\over z_{12}} \AYM{1,2,3,4} \ , \cr
}$$
and we used the following representation for the color-ordered tree-level SYM amplitude,
\eqn\fourSYM{
\AYM{1,2,3,4} = {1\over s_{12}}\langle V_{12}V_3V_4\rangle + {1\over s_{23}}\langle V_{1}V_{23}V_4\rangle \ .
}
Furthermore, using the explicit form $\cI^{(0)}_4 = |z_2|^{-\ap s_{12}} |1-z_2|^{-\ap s_{23}}$ of
the Koba--Nielsen factor at $\{z_1,z_3,z_4\}=\{0,1,\infty\}$, the integral in \fourtmp\ boils down to \dhokerS
\eqn\fourptInt{
\int d^2 z_2
z_2^{-{\ap\over 2} s_{12}-1} {\bar z}_2^{-{\ap\over 2}s_{12}-1}
(1-z_2)^{-{\ap\over 2} s_{23}}
(1-{\bar z}_2)^{-{\ap\over 2} s_{23}}
= 2\pi s_{23}^2 \halfap2  {\cal B}_0
}
with
\eqn\calBzero{
{\cal B}_0\equiv {\Gamma(-{\ap\over 2} s_{12})\Gamma(-{\ap\over 2} s_{13}) \Gamma(- {\ap\over 2} s_{14})
\over \Gamma(1+ {\ap\over2} s_{12})\Gamma(1+ {\ap\over2} s_{13})\Gamma(1+{\ap\over2} s_{14})}
= {3\over \s_3} + 2\zeta_3 + \zeta_5 \s_2 + {2\over 3}\zeta_3^2 \s_3 + \cdots \ .
}
Hence, the four-point amplitude \fourtmp\ is given by
\eqn\fourtree{
M^{(0)}_4 = (2\pi)^{10}\d^{10}(k)\halfap3 \kappa^4 e^{-2\l}\,2\pi \cK_4^{(0)}{\cal B}_0\,,
}
where (note $[\cK_4^{(0)}] = -8$)
\eqn\cKfourzero{
\cK_4^{(0)}\equiv | s_{12}s_{23}\AYM{1,2,3,4}|^2 
= |s_{23}\langle V_{12}V_3V_4\rangle + s_{12}\langle V_1 V_{23}V_4\rangle|^2\,.
}

\subsubsec The low-energy limit

From ${\cal B}_0 = (2/\ap)^3/(s_{12}s_{13}s_{14}) + \cdots$
and $\AYM{1,2,3,4} = \langle V_{12}V_3V_4\rangle/s_{12} + \langle V_1 V_{23}V_4\rangle/s_{23}$
the kinematic factor in the amplitude \fourtree\ becomes
\eqn\bcjNum{
-\halfap3 \cK_4^{(0)}{\cal B}_0 = -{|s_{12}s_{23}\AYM{1,2,3,4}|^2\over s_{12}s_{13}s_{14}} = {|\langle V_{12}V_3V_4\rangle |^2\over s_{12}}
+ {|\langle V_{31}V_2V_4\rangle |^2\over s_{13}} + {|\langle V_{23}V_1V_4\rangle |^2\over s_{23}}
}
where we used  $\langle V_{12}V_3V_4 \rangle + \langle V_{23}V_1V_4\rangle 
+ \langle V_{31}V_2V_4\rangle = 0$ \towards. Therefore the low-energy limit of \fourtree\ is given by
\eqn\fourlow{
M^{(0)}_4  = (2\pi)^{10}\d^{10}(k)\k^4 e^{-2\l}\, 2\pi \Big[
{|\langle V_{12}V_3V_4\rangle |^2\over s_{12}} + {|\langle V_{31}V_2V_4\rangle |^2\over s_{13}} +
{|\langle V_{23}V_1V_4\rangle |^2\over s_{23}}
\Big] + \cO(\ap^3) \ .
}

\subsec The five-point amplitude

According to the formula \treeFin, the five-point amplitude is given by
\eqn\fivetmp{
M^{(0)}_5 = (2\pi)^{10}\d^{10}(k)\halfap{-1}\! \kappa^5 e^{-2\l} \int \prod_{i=2}^{3} d^2 z_i |
\langle {\cal K}^{(0)}(z_2,z_{3})\rangle|^2 \cI^{(0)}_5\, ,
}
where
\eqnn\TODO
$$\eqalignno{
\langle {\cal K}^{(0)}(z_2,z_3)\rangle &= \langle \langle V_1 U_2(z_2)U_3(z_3) V_4 V_5\rangle \rangle &\TODO\cr
&= \halfap2 \Big[
{\langle V_{123}V_4V_5\rangle \over z_{12}z_{23}}
+ {\langle V_{12}V_{43} V_5\rangle \over z_{12}z_{43}}
+ {\langle V_{1}V_{432}V_5\rangle \over z_{43}z_{32}} + (2\leftrightarrow 3)\Big]\cr
&=\halfap2\, {s_{12}s_{34}\over z_{12}z_{34}}\AYM{1,2,3,4,5} + (2\leftrightarrow 3) \ .\cr
}$$
After inserting \TODO\ into \fivetmp, the $\alpha'$-expansion of the resulting integrals can be obtained through the KLT procedure \KLTref\ and arranged in the form \motivic
\eqnn\fivetree
\eqnn\KLT
$$\eqalignno{
M^{(0)}_5 &= (2\pi)^{10}\d^{10}(k)\halfap{}\; \kappa^5 e^{-2\l}(2\pi)^2 \cK_5^{(0)}
&\fivetree \cr
\cK_5^{(0)} &\equiv \tilde A^T_{54} \! \cdot \! S_0 \! \cdot \! \big[ 1 + 2\zeta_3\halfap3 M_3 + 2\zeta_5\halfap5 M_5 +2
\zeta_3^2 \halfap6 M_3^2 +  {\cal O}(\ap^7)\big] \! \cdot \! A_{45} \, ,&\KLT
}$$
where $\tilde A^T_{54}$ and $A_{45}$ are two-component vectors of SYM tree-amplitudes
\eqn\inKLT{
 \tilde A_{54}\equiv \pmatrix{\tilde A^{\rm YM}(1,2,3,5,4)\cr
 \tilde A^{\rm YM}(1,3,2,5,4)}\,,\quad
  A_{45} \equiv \pmatrix{ A^{\rm YM}(1,2,3,4,5)\cr A^{\rm YM}(1,3,2,4,5)} \, ,
}
and $S_0$ denotes the momentum kernel \BjerrumBohrHN, a convenient basis choice for the Mandelstam invariants in the KLT relations \KLTref
\eqn\momker{
S_0 \equiv  \pmatrix{s_{12}(s_{13}+s_{23}) & s_{12}s_{13}\cr
s_{12}s_{13} & s_{13}(s_{12}+s_{23})}\,.
}
The $2\times 2$ matrices $M_{2n+1}$ introduced in \motivic\ describe the momentum dependence of the $\alpha'$-corrections
and should not be confused with the amplitudes $M_n^{(g)}$. Their entries are degree $2n+1$ polynomials in
Mandelstam invariants, e.g. (see also \refs{\medina,\stieFive})
\eqnn\ininKLT
$$\eqalign{
 M_3 &\equiv \pmatrix{m_{11} & m_{12} \cr
                      m_{21} & m_{22}}
}\,,\quad\eqalign{
m_{12} &= - s_{13} s_{24} (s_1+s_2+s_3+s_4+s_5)\hskip88pt\ininKLT \cr
 m_{11} &= s_3 [ - s_1 (s_1+2s_2+s_3)+s_3s_4+s_4^2 ]+s_1s_5 (s_1+s_5)
}$$
with $m_{21} = m_{12} \big|_{2\leftrightarrow 3}$ and $m_{22} = m_{11} \big|_{2\leftrightarrow 3}$ as
well as $s_i \equiv s_{i,i+1}$ subject to $s_{5}= s_{15}$. Higher-order analogues such as $M_5$
relevant for the comparison with the two-loop five-point amplitude are available for download at the
website \WWWalpha. The overall coefficient of the five-point amplitude \fivetree\ will be verified by
factorization at the lowest order in $\alpha'$ in the appendix~\appUNI.

\newsec{One-loop closed-string amplitudes}

In this section the overall coefficients of the four- and five-point one-loop amplitudes are computed
using the conventions of section \secNorm, ensuring that the S-duality
analysis of section~\secSdual\ is unaffected by different conventions in the literature.
Although the
coefficient of the five-point amplitude can be derived from factorization (see appendix~\appUNI), its
computation from first principles as done in section~\subsecFive\ is novel
and validates the general method developed in \coefftwo.
For earlier references, see \GSoneloop\ for the original four-point derivation,
\refs{\sakai,\GreenOneLoopcoeff,\dhokerS,\coefftwo}
for discussions on its overall coefficient and
\refs{\tsuchiya,\montag,\StiebergerWK,\PierreFive,\RichardsJG,\oneloopNMPS,\oneloopbb} for related extensions.

\subsec The amplitude prescription

According to \amplitude, the $n$-point
closed-string one-loop prescription is
\eqn\amplitude{
M_n^{(1)} =
\half \k^n \!  \int
d^2\tau \int_{\Sigma_{n-1}}\!\!\!\!
\big|\langle\langle {\cal N}^{(1)}  (b,\mu)V^1(z_1) U^2(z_2) \cdots  U^n(z_n)\rangle\rangle \big|^2\, ,
}
where ${\cal N}^{(1)}$ is the genus-one instance of the zero-mode regulator \calNg\
and the $b$-ghost insertion \binsert\ reads
\eqn\binsert{
 (b,\mu) = {1\over 2\pi}\int d^2 y\, b(y) \mu(y)\, ,
}
where $\mu$ is the Beltrami differential for the modulus parameter $\tau$. In terms of the genus-one
period matrix $\Omega$, equation \mparam\ implies
\eqn\trouble{
\int d^2\tau \Big|  \int d^2 y\,  \om_1(y)\om_1(y) \mu(y)  \Big|^2 = \int d^2\Omega\,.
}
At genus one, there are $(16)_d$ zero-modes of $d_\a$ and $(11)_s$ zero-modes of $s^\a$.
Since there are no $s^\a$ variables in the vertex operators, and the term $s^\a \p\lb_\a$ from the
$b$-ghost \bghos\ does not contribute in absence of sources for $\wb^\a$,
the zero-modes of $s^\a$ are entirely saturated by the regulator through the factor
${\cal N}^{(1)}\rightarrow(s^1 d^1)^{11}$.
The remaining $(5)_d$ zero-modes must come from the $b$-ghost and the external vertices. 

There are two canonical $b$-ghost contributions to saturate the fermionic zero-modes 
of $d_\a$, with either one or two zero-modes. Expanding
$\Pi_m(y) = \Pi^1_m \om_1(y) +
\hat\Pi_m(y)$ and $d_\a(y) = d^1_\alpha \om_1(y) + \hat d_\alpha(y)$ where $\om_1(z)dz = dz$ is the genus-one
holomorphic one-form,
one can show
that amplitudes up to (and including) five-points receive zero-mode contributions from only two terms\foot{Terms
containing a single $N^{mn}$ zero-mode vanish upon integration over
$[dw][d\wb]$ \threeloop.} in the $b$-ghost \bghos
\eqn\bghostone{
\int d^2\tau |(b,\mu) |^2 = \halfap2
 {1\over (2\pi)^2}{1\over 192^2} \int d^2\Omega\,\big|B_{(2)} +  \Pi_m^1 B^m_{(1)} +\cdots\big|^2 \, ,
}
where the ellipsis represents terms relevant at $(n\geq 6)$ points and
\eqn\Bones{
B_{(2)} \equiv {1\over \llb^2}(\lb\g^{mnp} r)(d^1\g_{mnp}d^1),\qquad
B^m_{(1)} \equiv \invhalfap{}{96\over \llb}(\lb\g^m d^1) \ .
}
Since the vertex operators are independent of $w_\a, \wb^\a$ and $s^\a$, the integration over the
zero-modes $[dw^1][d\wb^1][ds^1]$
is readily performed using \Tds\ and yields
\eqn\cswintone{
\big|\int [ds^1][dw^1][d\wb^1]\, e^{-(d^1s^1) -(w^1\wb^1)}\big|^2 = \halfap4 {1\over(2\pi)^{16}2^2 Z_1^{22}}
\Big|{(\e\cdot T\cdot d^1)\over(11!\,5!)} \Big|^2 \ .
}
Defining
\eqn\Dtwelvedef{
D_{(13)} \equiv \int [dd^1] {(\e\cdot T\cdot d^1)\over (11!\,5!)} B_{(2)},\quad
D^m_{(12)} \equiv \int [dd^1] {(\e\cdot T\cdot d^1)\over (11!\,5!)} B^m_{(1)}\,,
}
as a special case of \definitionD, the amplitude \amplitude\ becomes
\eqn\ampgen{
M^{(1)}_n =
\halfap6\! {\k^n\over (2\pi)^{18}2^{15}\,3^2}
 \int {d^2\Omega\over Z_1^{22}}
\int_{\Sigma_{n-1}}\!\!\!|\langle \langle {\cal K}_{[d]}^{(1)}(z_2, \ldots,z_n) \rangle \rangle_{(3,1)}|^2\,
\Big \langle\prod_{j=1}^n \!e^{k^j\cdot x^j} \Big \rangle \ .
}
The subscript $[d]$ of the kinematic factor
\eqn\calKonedef{
\cK^{(1)}_{[d]}(z_2, \ldots,z_n) \equiv (D_{(13)}
+ \Pi^m_1 D^m_{(12)}+\cdots) V_1 U_2(z_2) \cdots U_n(z_n)
}
emphasizes the remaining integration over the $d_\a$ zero-modes. The ellipsis along with
$\Pi^m_1 D^m_{(12)}$ refers to $b$-ghost contributions 
which do not affect $(n \leq 5)$-point amplitudes.

\subsubsec Scalar and vector building blocks at genus one

The integration over the zero-mode $d^1_\a$ in \calKonedef\ can be done using \Tds\ and gives
\eqnn\ddOne
$$\eqalignno{
D_{(13)}V_A  (d^1 W_B) (d^1 W_C) (d^1 W_D) &=  96\, c_d T_{A|B,C,D}(\l,\lb),&\ddOne\cr
D^m_{(12)}V_A (d^1 W_B) (d^1 W_C) (d^1 W_D)(d^1 W_E) &=  96\, c_d \invhalfap{} S^m_{A|B,C,D,E}(\l,\lb)\, ,
}$$
where
\eqnn\TodNMPSdef
$$\eqalignno{
T_{A|B,C,D}(\l,\lb) &\equiv {(\lb\g_{mnp}r)\over \llb^2} V_A(\l\g^m W_{B})(\l\g^n W_C)(\l\g^p W_D), &\TodNMPSdef\cr
S^m_{A|B,C,D,E}(\l,\lb) &\equiv {(\lb\g^{m}\g^r\l)\over \llb} V_A (\l\g^s W_B)(\l\g^t W_C)(W_D \g_{rst} W_E) \cr
}$$
with multiparticle labels $A,B,\ldots$ (see section~\secMP).

At this stage the
Theorem~1 from \threeloop\ can be used to factorize $\llb$ from the expressions in \TodNMPSdef. For
a general kinematic factor one then defines $K_{A|B, \ldots}(\l,\l) = \llb^p K_{A|B, \ldots}$ for some
power $p$ as the result of this procedure.
Doing this for \TodNMPSdef\ leads to
\eqnn\TheoTs
$$\eqalignno{
\langle T_{A|B,C,D}(\l,\lb)\rangle_{(3,1)} &= \langle T_{A|B,C,D}\rangle_{(2,1)}, &\TheoTs\cr
\langle S^m_{A|B,C,D,E}(\l,\lb)\rangle_{(3,1)} &= \langle S^m_{A|B,C,D,E}\rangle_{(3,1)} = 10 \langle
S^m_{A|B,C,D,E}\rangle_{(2,1)}\, ,
}$$
where symmetry of $T_{A|B,C,D}$ and $S^m_{A|B,C,D,E}$ in $(B,C,D)$ and $(B,C,D,E)$, respectively, is
inherited from \ddOne. Note that any appearance of $S^m_{A|B,C,D,E}$ in $(n\geq 5)$-point one-loop amplitudes occurs in the combination\foot{Writing the term $A^m_1 T_{2|3,4,5}$ is an abuse of notation
since when computing its
component expansion the variables $r_\a$ in the definition of $T_{2|3,4,5}(\l,\lb)$ become covariant
derivatives $D_\a$ (see \PSanomaly) and must also act upon the superfield $A^m_1$.}
\eqn\TmOnedef{
T^m_{A|B,C,D,E} \equiv A^m_B T_{A|C,D,E} + A^m_C T_{A|B,D,E} + A^m_D T_{A|B,C,E} + A^m_E T_{A|B,C,D}
+ 10  S^m_{A|B,C,D,E}\,,
}
where the factor of $10$ is due to the conversion from $\langle \ldots\rangle_{(3,1)} = 10\langle
\ldots\rangle_{(2,1)}$  in \TheoTs.

\subsec{The four-point amplitude}

According to the formula \ampgen, the four-point amplitude is given by
\eqn\amptmp{
M^{(1)}_4 =
\halfap6{\k^4\over (2\pi)^{18}2^{15}\,3^2}
 \int {d^2\Omega\over Z_1^{22}}
\int_{\Sigma_3} \;  |\langle \langle {\cal K}^{(1)}_{[d]}(z_2,z_3,z_4) \rangle \rangle_{(3,1)}|^2\Big \langle \KN4 \Big\rangle \,.
}
It is easy to see that $D_{(13)}$ is the only non-vanishing contribution from the $b$-ghost
since the external vertices cannot provide four $d_\a$ zero-modes to saturate the $D^m_{(12)}$
integral \NMPS. The integration over $[dd^1]$ is readily performed via \ddOne\ followed by \TheoTs,
\eqn\calKfour{
\langle \langle \cK^{(1)}_{[d]}(z_2,z_3,z_4) \rangle \rangle_{(3,1)}
= 96 c_d \halfap3 \langle T_{1|2,3,4}\rangle_{(2,1)} \ .
}
Note that the right-hand side is independent on the
vertex insertion points $z_2$, $z_3$ and $z_4$ because only the zero-modes entered the
computation.
A straightforward application of \Nthreezero\ then implies
\eqnn\fourtmp
$$\eqalignno{
M^{(1)}_4 &=
(2\pi)^{10}\d^{10}(k) \halfap3 {\k^4\over 2^{14}\,5^2\pi^2}|\langle T_{1|2,3,4}\rangle|^2
 \int {d^2\Omega\over (\Im\Omega)^5}\int_{\Sigma_3} \! \cI_4^{(1)}, &\fourtmp\cr
&= (2\pi)^{10}\d^{10}(k)\halfap3 {\k^4 \over 2^8 \pi^2}
\cK_4^{(1)}
 \int {d^2\Omega\over (\Im\Omega)^5}
\int_{\Sigma_3} \! \cI_4^{(1)},
}$$
where in the second line we used \tese\ (note $[\cK_4^{(1)}] = -8$)
\eqn\forty{
\langle T_{1|2,3,4}\rangle = 40\langle V_1 T_{2,3,4}\rangle,\quad
\cK_4^{(1)}\equiv |\langle V_1 T_{2,3,4}\rangle|^2\,.
}
Note that the tree-level \cKfourzero\ and one-loop \forty\
kinematic factors are related by \mafraids
\eqn\TreeOne{
\cK_4^{(1)} = \cK_4^{(0)}\,,
}
a well-known result first obtained by Green and Schwarz \GSoneloop.

\subsubsec The $\ap$-expansion of the four-point amplitude

The $\ap$-expansion of the four-point amplitude\foot{In addition to the analytic momentum dependence
shown in \calBone, threshold singularities arise from the integration region where $\Im
\Omega\rightarrow \infty$. A careful treatment of these non-analytic terms can be found in
\GreenOneLoopcoeff.} has been extensively studied in a series of papers \GreenOneLoopcoeff, where the
subleading term in
\eqn\calBone{
\int {d^2\Omega\over (\Im\Omega)^5}
\int_{\Sigma_3} \! \cI_4^{(1)} = {2^4\pi\over 3}\big(1 + {\zeta_3\over 3}\sigma_3 + \cdots \big)
}
signals the absence of $D^4R^4$ interactions at one-loop in ten dimensions. Therefore, plugging the above
result in the four-point amplitude \fourtmp\ leads to
\eqn\leadingFourOne{
M^{(1)}_4
= (2\pi)^{10}\d^{10}(k)\halfap3 {\k^4 \over 2^4\,3 \pi}
\cK_4^{(1)} + \cO(\ap^6)\,.
}

\subsec The five-point amplitude
\par\subseclab\subsecFive

\noindent Using the general result \ampgen\ the five-point amplitude \amplitude\ becomes
\eqn\amptmp{
M^{(1)}_5 =
{\k^5\over (2\pi)^{18}2^{15}\,3^2}\halfap6
 \int {d^2\Omega\over Z_1^{22}}
\int_{\Sigma_4} \; |\langle \langle{\cal K}^{(1)}_{[d]}(z_2, \ldots,z_5)\rangle \rangle_{(3,1)}|^2\Big \langle
\prod_{j=1}^5 \!e^{k^j\cdot x^j}  \Big \rangle  \, ,
}
where
\eqn\calKoneFive{
\cK^{(1)}_{[d]}(z_2, \ldots,z_5) = (D_{(13)}
+ \Pi_m^1 D^m_{(12)}) V_1 U_2(z_2) \cdots U_5(z_5) \ .
}
The $[dd^1]$ integration with the operators of \Dtwelvedef\ picks up
the terms with four and three $d_\alpha$ zero-modes from the
vertices, respectively. Using the multiparticle superfields of \refs{\EOMBBs,\HighSYM} one arrives at
\eqnn\fourds
$$\eqalignno{
V_1 U_2 U_3 U_4 U_5\Big|_{d^4} &=  \halfap4 V_1 (d^1 W_2)(d^1 W_3)(d^1 W_4)(d^1 W_5)&\fourds\cr
V_1 U_2 U_3 U_4 U_5\Big|_{d^3} &= \halfap4 V_{12}(d^1W_3)(d^1W_4)(d^1W_5)\eta_{12} + (2|2,3,4,5)\cr
& + \halfap4 V_1 (d^1 W_{23})(d^1 W_{4})(d^1 W_{5})\eta_{23} + (2,3|2,3,4,5)\cr
& + \halfap3 \Pi^1_m V_1 A_2^m (d^1 W_3)(d^1 W_4)(d^1 W_5) + (2|2,3,4,5)\,,\cr
}$$
where the notation $(A_1,A_2,\ldots, A_p \,|\, A_1,A_2,\ldots,A_n)$ instructs to sum over all possible
ways to choose $p$ elements $A_1,A_2,\ldots ,A_p$ from the set $\{A_1,{\ldots} ,A_n\}$, for a total of
${n\choose p}$ terms. We have $\om_1(z)dz = dz$ for the genus-one surface, and $\eta_{ij} ={1\over z_{ij}}+{\cal O}(z_{ij})$ defined by
\defeta\ accounts for the singularity from the OPEs among
vertex operators. According to \VUOPE\ and \UUOPE, they introduce multiparticle
superfields $W_{23}^\alpha$ and $V_{12}$ defined in \twopart\ and \Vtwo, respectively.

The integration over the zero-modes of $d_\a$ uses
the formulas \ddOne\ and \TheoTs\ to yield
\eqn\calKs{
\langle \langle \cK^{(1)}_{[d]}(z_2, \ldots,z_5)\rangle \rangle_{(3,1)} = 96\, c_d \halfap3 \langle \cK^{(1)}(z_2,
\ldots,z_5)\rangle_{(2,1)}\,,
}
where
\eqnn\Ktwo
$$\eqalignno{
 \cK^{(1)}(z_2, \ldots,z_5) & \equiv
\halfap{}\bigl[\eta_{12}  T_{12|3,4,5}  + (2|2,3,4,5)\bigr] &\Ktwo\cr
&+  \halfap{}\bigl[\eta_{23} T_{1|23,4,5} + (2,3|2,3,4,5)\bigr]
+  \Pi^1_m T^m_{1|2,3,4,5}\,,
}$$
see \TmOnedef\ for the definition of $T^m_{1|2,3,4,5}$. Upon discarding $\Pi^1_m\rightarrow 0$ and
adjoining the Koba-Nielsen factor, this is precisely the open-string correlation function for
the five-point pure spinor one-loop amplitude \refs{\oneloopNMPS,\oneloopbb} (for the RNS derivation, see \refs{\tsuchiya,\StiebergerWK}).

Therefore, the closed-string amplitude \amptmp\ becomes
\eqnn\AmpliFinal
$$\eqalignno{
 M^{(1)}_5 &=
  {\k^5 \over  2^{7} \pi^2} \halfap{4}
  \int {d^2\Omega\over Z_1^{-10}}  \int_{\Sigma_4}
  \big|\langle \cK^{(1)}(z_2, \ldots,z_5)\rangle_{(2,1)}\big|^2 \Big \langle \KN5 \Big \rangle\cr
&=(2\pi)^{10}\d^{10}(k){\k^5 \over  2^{14}\,5^2 \pi^2} \halfap{3}
  \int {d^2\Omega\over (\Im\Omega)^5}  \int_{\Sigma_4}
  \big|\langle \cK^{(1)}(z_2, \ldots,z_5)\rangle \big|^2 \cI^{(1)}_5 \ , &\AmpliFinal\cr
}$$
where we used $Z_1^{-10} = (2\Im\Omega)^5$ and the identity \Nthreezero\ on the second line.
Integration by parts identities \refs{\RichardsJG, \FiveSdual} allow to express
\AmpliFinal\ in terms of 37 basis integrals with BRST-invariant kinematic numerators. The $\ap$-expansion of these integrals was analyzed in \refs{\RichardsJG, \FiveSdual} and confirms the absence of $D^2 R^5$ interactions at one-loop in ten dimensions.

\subsubsec The leading-order contribution

\noindent The low-energy behavior of the $\Sigma_4$ integral over $\big|\langle \cK^{(1)}(z_2,
\ldots,z_5)\rangle \big|^2 \cI^{(1)}_5 $ in \AmpliFinal\ is governed by two kinds of contributions \refs{\RichardsJG,
\FiveSdual}:
\smallskip
\item{(i)} zero-mode contractions $\Pi^1_m \bar \Pi^1_n \rightarrow - \eta_{mn} \big(\! {\ap \over 2}\! \big){\pi\over \Im\Omega} $
following \LRcontract\ at $g=1$

\item{(ii)} kinematic poles\foot{Strictly speaking, the identity \etares\ is valid under integration over one
of $z_1,z_2$ and results from the behavior of the Koba-Nielsen factor ${\cal I}^{(g)}_n \sim
|z_{12}|^{- \ap s_{ij}}$ as $z_1\rightarrow z_2$:
$$
\int d^2 z_2 \eta_{12} \bar \eta_{12} {\cal I}_n^{(g)} =
4\pi \int |z_{12}| d |z_{12}| {1 \over |z_{12}|^2} |z_{12}|^{-\ap s_{12}} + {\cal O}(\ap^0) = -{4\pi\over \ap s_{12}}
+ {\cal O}(\ap^0)
$$
This only depends on the local properties of the worldsheet and therefore holds at any genus.} from the
residue of the pole $\eta_{12} \bar \eta_{12}\sim |z_{12}|^{-2}$
\eqn\etares{\eta_{12} \bar \eta_{12} {\cal I}^{(g)}_n = -\invhalfap{} {2\pi\over s_{12}} \delta^2(z_1-z_2) {\cal I}^{(g)}_n + {\cal O}(\ap^0)\ .
}
\smallskip
\noindent Non-diagonal products of Green functions such as $\eta_{12} \bar \eta_{13}$ or $\eta_{12} \bar
\eta_{34}$ do not contribute to the leading order in $\ap$. 
Hence, we have
\eqn\effone{
\int_{\Sigma_4}
  \big|\langle \cK^{(1)}(z_2, \ldots,z_5)\rangle \big|^2 \cI^{(1)}_5 =
  -\halfap{}{\pi\over \Im\Omega} \cK^{(1)}_5 \int_{\Sigma_4}{} + {\cal O}(\ap^2) \ ,
}
where the kinematic factor $\cK^{(1)}_5$ is defined by (note $[\cK_5^{(1)}] = -8$)
\eqn\calKFiveOne
{
\cK^{(1)}_5  \equiv
\Big[{|\langle T_{12|3,4,5}\rangle |^2\over s_{12}} + (2|2,3,4,5)\Big]
+ \Big[{|\langle T_{1|23,4,5}\rangle |^2\over s_{23}} + (2,3|2,3,4,5)\Big]
+ |\langle T^m_{1|2,3,4,5}\rangle|^2\ ,
}
and the integration over $\Sigma_4$ gives $\int_{\Sigma_4} = 2^4\Im\Omega^4$. This leads
to the following result for the one-loop five-point amplitude (recall that $\int d\mu_1 = 2\pi/3$)
\eqnn\LowEoneFinal
$$\eqalignno{
M^{(1)}_5 &=
(2\pi)^{10}\d^{10}(k)\halfap{4} {\k^5 \over  2^{9}\, 5^2\,3}\, \cK^{(1)}_5
 + \cO(\ap^5)\,, &\LowEoneFinal
}$$
In the appendix~\appUNI\ the overall coefficient in \LowEoneFinal\ will be
validated by factorization.

\subsubsec Components in type IIB and type IIA

The type IIB components of the kinematic factor \calKFiveOne\
are related to the first $\alpha'$-correction of the five-point tree-level amplitude \fivetree\ and \KLT\ \PSS:
\eqn\CompFive{
\cK^{(1)}_5 \Big|_{\rm IIB} = 2^5\,5^2 \halfap{-3}\; \cK^{(0)}_5\Big|_{\zeta_3} \times \cases{\ \ \, 1 \,  \ : \ {\rm five} \  {\rm gravitons} \cr -{1\over 3} \ : \ {\rm four} \ {\rm gravitons}, \ {\rm one} \ {\rm dilaton}}
}
The relative factor between the tree-level and one-loop amplitudes at order $\ap^4$ turns out to depend
on the charges of the external states under the R-symmetry of type IIB supergravity, as has already
been observed in \FiveSdual. Components with the same R-symmetry violation as four gravitons and one
dilaton give rise to an additional relative factor of $-{1\over 3}$. This will be explained in
section~\secDilatonS\
from an S-duality point of view. Since R-symmetry violating four-point amplitudes vanish
\Rviolating, the five-point amplitudes in this work provide the simplest context to study the
S-duality properties of interactions with R-charge. Also, five-point amplitudes that violate R-charge
by more units than caused by a single dilaton insertion vanish at any loop-order.

Type IIA components of the five-point low-energy limit \LowEoneFinal\ cannot
be expressed in terms of $A^{\rm YM}$ bilinears. Instead, we have
\eqn\CompFiveA{
\cK^{(1)}_5 \Big|_{\rm IIA}^{5 \ {\rm gravitons}} = 2^5\,5^2 \halfap{-3}\; \Big[ \cK^{(0)}_5\Big|_{\zeta_3} 
- \big|\epsilon^{me^1k^2e^2k^3e^3k^4e^4k^5e^5}\big|^2 \Big] \, ,
}
where the notation $\epsilon^{me^1k^2e^2k^3e^3k^4e^4k^5e^5}\equiv
\epsilon_{10}^{mn p_2q_2\ldots p_5 q_5} e_n^1 k_{p_2}^2 e_{q_2}^2\ldots k_{p_5}^5 e_{q_5}^5$ has been
used and the free vector index $m$ is contracted between the left- and right-moving factors in the
holomorphic square. The parity-violating type IIA component with a $B$-field and four gravitons has been evaluated in \FiveSdual.

Upon insertion into \LowEoneFinal, the kinematic factors \CompFive\ and \CompFiveA\ give rise to the following
low-energy limits for the five-graviton amplitudes:
\eqnn\LEtmpfinaloneB
$$\eqalignno{
M_5^{(1)} \big|^{\ap^4}_{{\rm IIB} \ {\rm gravitons}}
&= (2\pi)^{10}\d^{10}(k)\halfap{} {\k^5 \over  2^{4} 3 } 
 \cK_5^{(0)}\Big|_{\zeta_3}
 &\LEtmpfinaloneB
 \cr
 M_5^{(1)} \big|^{\ap^4}_{{\rm IIA}\ {\rm gravitons}}
&= (2\pi)^{10}\d^{10}(k)\halfap{} {\k^5 \over  2^{4} 3 }
\Big[ \cK_5^{(0)}\Big|_{\zeta_3}
-  \big|\epsilon^{me^1k^2e^2k^3e^3k^4e^4k^5e^5}\big|^2 \Big] \, .
}$$
According to \CompFive, the R-symmetry violating type IIB components (e.g. four gravitons and one dilaton) carry an extra factor of $-{1\over3}$.

\newsec Two-loop closed-string amplitudes

In this section we compute the low-energy limit of the
two-loop five-point amplitude including its overall coefficient from first principles. This includes a recomputation of the four-point amplitude
using the conventions of section~\secNorm.
For previous two-loop four-point results see \refs{\dhokerVI,\twoloop,\twolooptwo,\dhokerS,\coefftwo}.

\subsec The amplitude prescription

The $n$-point two-loop amplitude prescription \amplitude\ is given by
\eqn\amplitudetwo{
M_n^{(2)} = \half
\k^n e^{2\l}  \int \prod_{j=1}^3 d^2\tau_j \int_{\Sigma_n}\big|
\langle\langle {\cal N}^{(2)}  (b,\mu_j) U^1(z_1)\cdots U^n(z_n)\rangle \rangle \big|^2 \, ,
}
where the zero-mode regulator ${\cal N}^{(2)}$ is defined in \calNg, and the $b$-ghost
insertion was specified in \binsert. At genus two, there are $(16,16)_d$ zero-modes of $d_\a$ and $(11,11)_s$ zero-modes of $s^\a$. The latter are entirely saturated by the regulator through the factor
${\cal N}^{(2)}\rightarrow(s^1 d^1)^{11}(s^2 d^2)^{11}$, see the discussion below \trouble.

In presence of five vertex operators, it is easy to see that the total number of $d_\a$ zero-modes from the $b$-ghosts
can be distributed as $(p,q)$ such that $p+q$ is either $5$ or $6$. These two contributions
can be separately computed using the zero-mode expansion
$$
(d\g_{mnp}d)(z) \rightarrow (d^1\g_{mnp}d^1)\om_{1}(z)\om_1(z) + 2 (d^1\g_{mnp}d^2)\om_1 (z)\om_2(z) +
(d^2\g_{mnp}d^2)\om_{2}(z)\om_2(z)
$$
and the general formulas \mparam\ 
as follows
\eqn\bghostds{
\int \prod_{j=1}^3 d^2\tau_j \Big|(b,\mu_j)\Big|^2
=\halfap6 {1\over (2\pi)^6 192^6} \int d^2\Omega\,
 \big|B_{(3,3)}
+ \big(\Pi_m^1 B^m_{(2,3)} + \Pi_m^2 B^m_{(3,2)} \big)+\cdots \big|^2 \,.
}
The shorthands for different $b$-ghost contributions are defined by\foot{The $(5,5)_d$ zero-modes from
the $b$-ghosts and the vertices can in principle be saturated by a $b$-ghost contribution
$B_{(1,4)}^m \sim (d^2 d^2)(d^1 d^2) (\Pi^m d^2)$. However, the integration over the $b$-ghost
insertions via \mparam\ yields a $\tau_j$ integrand $\sim \epsilon_{i_1i_2i_3} {\delta
\Omega_{22} \over \delta \tau_{i_1}}{\delta \Omega_{12} \over \delta \tau_{i_2}} \big( \Pi^m_1{\delta
\Omega_{12} \over \delta \tau_{i_3}}+ \Pi^m_2 {\delta \Omega_{22} \over \delta \tau_{i_3}} \big)$ whose summands
do not depend on all entries of the period matrix and which vanish upon contraction with the antisymmetric
$\epsilon_{i_1i_2i_3}$. The same mechanism suppresses $\Pi_m^2 B^m_{(2,3)}$ and $\Pi_m^1 B^m_{(3,2)}$ from \bghostds.}
\eqnn\Bzeros
$$\eqalignno{
B_{(3,3)} &\equiv {1\over \llb^6}\big[ 2(\lb r d^1d^1)(\lb r d^1d^2)(\lb r d^2d^2)\big] &\Bzeros\cr
B^m_{(2,3)} &\equiv \invhalfap{}{96\over \llb^5}\big[ 2(\lb \g^m d^1) (\lb r d^1d^2)  (\lb r d^2d^2)
- (\lb \g^m d^2) (\lb r d^1d^1) (\lb r d^2d^2)\big] \cr
B^m_{(3,2)} &\equiv \invhalfap{}{96\over \llb^5}\big[ 2 (\lb \g^m d^2) (\lb r d^1d^1) (\lb r d^1d^2)
- (\lb \g^m d^1) (\lb r d^1d^1) (\lb r d^2d^2)\big] \cr
}$$
with the convention that $(\lb r d^Id^J) \equiv (\lb\g^{mnp} r)(d^I\g_{mnp}d^J)$. Note that $B_{(3,3)}\rightarrow - B_{(3,3)}$ and $B_{(2,3)}^m \leftrightarrow -B_{(3,2)}^m$ under the
interchange of zero-mode labels $d^1 \leftrightarrow d^2$. As indicated by the ellipsis in \bghostds, two-loop amplitudes involving $n\geq 6$ closed-string states allow for additional $b$-ghost contributions with fewer zero-modes of $d_\alpha$.

Since the vertex operators are independent of $w_\a^I$,
$\wb_I^\a$ and $s_I^\a$, the integration over their zero-modes can be performed at an early stage
using \Tds,
\eqn\FirstIntTwo{
\Bigl|\int \prod_{I=1}^2 [dw^I][d\wb^I][ds^I]\, e^{-(w^I\wb^I) - (d^I s^I)}\Bigr|^2 =
\halfap{8}\!{1\over (2\pi)^{32} 2^4 Z_2^{22}}\Bigl|\prod_{I=1}^2 {(\e\cdot T\cdot d^I)\over
(11!\,5!)}\Bigr|^2 \ .
}
The tensor structure $(\e\cdot T\cdot d^I)$ is captured by the operators $D_{(14,14)}$ and $D^m_{(14,13)}$ defined in \definitionD.
They allow to rewrite the amplitude \amplitudetwo\ as
\eqn\twogeneral{
M_n^{(2)} =
\halfap{14}\!\!\!{\k^n e^{2\l}\over (2\pi)^{38}2^5 192^6}  \int {d^2\Omega\over Z_2^{22}} \int_{\Sigma_n}
\big|\langle \langle \cK^{(2)}_{[d]}(z_1, \ldots,z_n) \rangle \rangle_{(3,2)}\big|^2\, \Big \langle \prod_{j=1}^n e^{k^j\cdot x^j}\Big \rangle \, ,
}
where
\eqn\caltwodef{
\cK^{(2)}_{[d]}(z_1, \ldots,z_n) \equiv \big( D_{(14,14)}
+ \Pi_m^1 D^m_{(13,14)} + \Pi_m^2 D^m_{(14,13)}+\cdots \big)
U^1(z_1)\cdots U^n(z_n)\,.
}
The ellipsis along with $\Pi_m^2 D^m_{(14,13)}$ accounts for $b$-ghost zero-mode contributions
which drop out from the subsequent four- and five-point computations.

\subsubsec Scalar and vector building blocks at genus two

We shall now evaluate \caltwodef\ on the part of the vertex operators which contribute zero-modes $d^1,d^2$. One can show that
\eqnn\DIntsOne
$$\eqalignno{
D_{(14,14)} (d^1W_{A})(d^1W_{B})(d^2W_C)(d^2W_D) &= 96^2 c_d^2\, T_{A,B|C,D}(\l,\lb)
&\DIntsOne\cr
D^m_{(13,14)}(d^1 W_A)(d^1 W_B)(d^1 W_C)(d^2 W_D)(d^2 W_E) &=  \invhalfap{}\, 96^2 c_d^2\, S^m_{A,B,C|D,E}(\l,\lb)
\cr
D^m_{(14,13)}(d^2 W_A)(d^2 W_B)(d^2 W_C)(d^1 W_D)(d^1 W_E) &=  \invhalfap{}\, 96^2 c_d^2\, S^m_{A,B,C|D,E}(\l,\lb) 
}$$
with multiparticle labels $A,B,\ldots$ and scalar building block
\eqnn\MDs
$$\eqalignno{
T_{A,B|C,D}(\l,\lb) &\equiv {2\over \llb^6} (\lb\g_{m_1n_1p_1}r)(\lb\g_{def}r)(\lb\g_{m_2n_2p_2}r)(\l\g^{m_1defm_2}\l) &\MDs\cr
 &\quad\times (\l\g^{n_1} W_{A})(\l\g^{p_1} W_B)(\l\g^{n_2} W_C)(\l\g^{p_2}W_D) \ .
}$$
The two zero-mode patterns $(\lb \g^m d^1) (\lb r d^1d^2)  (\lb r d^2d^2)$ and $(\lb \g^m d^2) (\lb r d^1d^1) (\lb r d^2d^2)$ in the $b$-ghost contribution $B^m_{(3,2)}$ given by \Bzeros\ lead to distinct tensor structures $S^{(1)\,m}_{A,B,C|D,E}$ and $S^{(2)\,m}_{A,B,C|D,E}$ such that
\eqn\Smtwodef{
S^m_{A,B,C|D,E} \equiv S^{(1)\,m}_{A,B,C|D,E} + S^{(2)\,m}_{A,B,C|D,E}
} 
with
\eqnn\Sone
\eqnn\Stwo
$$\eqalignno{
S^{(1)\,m}_{A,B,C|D,E}(\l,\lb) &\equiv -{2\over \llb^5}\,(\lb\g^m \g^{a_1}\l)(\lb\g_{m_1n_1p_1} r) (\lb\g_{m_2n_2p_2} r) (\l\g^{a_2m_1n_1p_1m_2}\l)
\cr
&\quad\times  (W_A\g_{a_1a_2a_3}W_B) (\l\g^{a_3} W_C) (\l\g^{n_2} W_D) (\l\g^{p_2} W_E)\,, &\Sone\cr
S^{(2)\,m}_{A,B,C|D,E}(\l,\lb) &\equiv {96\over \llb^5} \,(\lb\g^m \g^{b_1}\l)(\lb\g_{a_1a_2a_3} r) (\lb\g_{b_1b_2b_3} r)\cr
&\quad\times(\l\g^{a_1} W_A) (\l\g^{a_2} W_B) (\l\g^{a_3} W_C) (\l\g^{b_2} W_D) (\l\g^{b_3} W_E)\,. &\Stwo\cr
}$$
Note that the integrals of $D_{(13,14)}^m$ and $D_{(14,13)}^m$ give rise to the same kinematic
structure $S^m_{A,B,C|D,E}$ because
$D^m_{(13,14)} \leftrightarrow D^m_{(14,13)}$ under the interchange of zero-modes
$d^1_\a \leftrightarrow d^2_\a$.

\subsubsec Kinematic symmetry properties at genus two

The above definitions in \DIntsOne\ manifest the symmetry properties
\eqnn\fivesymmC
$$\eqalignno{
T_{A,B|C,D}(\l,\lb) &= T_{(A,B)|(C,D)}(\l,\lb) \ ,
\ \ \ \
S^{(j)\,m}_{A,B,C|D,E}(\l,\lb) = S^{(j)\,m}_{(A,B,C)|(D,E)}(\l,\lb)   &\fivesymmC
}$$
with $j=1,2$. As demonstrated in the appendix \appJac, gamma-matrix manipulations and the pure
spinor constraint imply that the kinematic factor \MDs\ can be rewritten as
\eqn\TijFivedef{
T_{A,B|C,D}(\l,\lb) =
-{192\over \llb^4} (\lb\g_{a mn}r)(r\g_{a pq}r)(\l\g^{m} W_{A})(\l\g^{n} W_B)(\l\g^{p} W_C)(\l\g^{q}W_D)
}
and satisfies
the Jacobi identity,
\eqn\jacId{
T_{A,B|C,D}(\l,\lb) + T_{A,D|B,C}(\l,\lb) + T_{A,C|D,B}(\l,\lb) = 0\,.
}
The symmetry \jacId\ assembles the holomorphic one-forms in the antisymmetric combinations $\Delta_{ij}
= \e^{IJ}\omega_I(z_i)\omega_J(z_j)$,
\eqnn\whyjacobi
$$\eqalignno{
&D_{(14,14)} (dW_A)(z_A) (dW_B)(z_B) (dW_C)(z_C) (dW_D)(z_D) \cr
 &=96^2 c_d^2 \Big( T_{A,B|C,D}(\l,\lb) \big[ \om_1(z_A) \om_1(z_B)\om_2(z_C) \om_2(z_D) + \om_2(z_A) \om_2(z_B)\om_1(z_C) \om_1(z_D)\big] \cr
 & \ \ + T_{A,C|B,D}(\l,\lb) \big[ \om_1(z_A) \om_1(z_C)\om_2(z_B) \om_2(z_D) + \om_2(z_A) \om_2(z_C)\om_1(z_B) \om_1(z_D)\big] \cr
 &\ \ + T_{A,D|B,C}(\l,\lb) \big[ \om_1(z_A) \om_1(z_D)\om_2(z_B) \om_2(z_C) + \om_2(z_A) \om_2(z_D)\om_1(z_B) \om_1(z_C)\big] \Big) \cr
 &=-96^2 c_d^2\big[ T_{A,B|C,D}(\l,\lb) \Delta_{DA} \Delta_{BC} +  T_{D,A|B,C}(\l,\lb) \Delta_{AB} \Delta_{CD}  \big] \ .
 &\whyjacobi
}$$
As will become clear later, the appearance of $\Delta_{ij}$ in \whyjacobi\ is a crucial requirement for modular
invariance of the amplitude.

Furthermore, it is shown in appendix \appJac\ that $S^{(1)m}_{A,B,C|D,E}$ can be eliminated in favor of $S^{(2)m}_{A,B,C|D,E}$ to yield
\eqn\Smfinal{
S^{m}_{A,B,C|D,E}(\l,\lb) = 2S^{(2)\,m}_{A,B,C|D,E}(\l,\lb) + S^{(2)\,m}_{A,D,E|B,C}(\l,\lb)
+ S^{(2)\,m}_{B,D,E|A,C}(\l,\lb) + S^{(2)\,m}_{C,D,E|A,B}(\l,\lb) .
}
Together with the ten-term identity\foot{The identity \tenterm\ was checked to hold for its bosonic
(gluon) components \PSS, and it is believed to hold at the superfield level using similar
manipulations seen in the appendix \appJac.}
\eqn\tenterm{
S^{(2)\,m}_{A,B,C|D,E}(\l,\lb) +
(D,E|A,B,C,D,E)=0\,,
}
one can show that \Smfinal\ implies a vector generalization of the scalar
Jacobi identity \jacId \eqnn\fivesymmE
$$\eqalignno{
S^{m}_{A,B,C|D,E}(\l,\lb)&=  
S^{m}_{C,D,E|A,B}(\l,\lb)+S^{m}_{B,D,E|A,C}(\l,\lb)+S^{m}_{A,D,E|B,C}(\l,\lb) \ . &\fivesymmE
}$$
This is instrumental to identify $\Delta_{ij}$ in the following permutation sum:
\eqnn\whyjac
$$\eqalignno{
&(\Pi_m^1 D^m_{(13,14)}+\Pi_m^2 D^m_{(14,13)}) (dW_A)(z_A) (dW_B)(z_B) (dW_C)(z_C) (dW_D)(z_D) (dW_E)(z_E)
\cr
&= 96^2 c_d^2 \Big( {2\over \alpha'} \Big) S^m_{A,B,C|D,E}(\l,\lb)   \big[ \Pi_m^1 \om_1(z_A) \om_1(z_B)\om_1(z_C) \om_2(z_D)\om_2(z_E) \cr
& \ \ \ \ \ \ \ \ \ \ \ \ \ \ \ +\Pi_m^2 \om_2(z_A) \om_2(z_B)\om_2(z_C) \om_1(z_D)\om_1(z_E) \big]  + (D,E|A,B,C,D,E) \cr
&= 96^2 c_d^2 \Big( {2\over \alpha'} \Big) S^m_{A,B,C|D,E}(\l,\lb)   \sum_{I=1}^2 \Delta_{EA} \omega_I(z_B) \Delta_{CD} \Pi_m^I + {\rm cyc}(A,B,C,D,E) \ .
&\whyjac
}$$
Applying the Theorem~1 of \threeloop\ to the expressions \MDs\ and \Smtwodef\
\eqn\Tcompdefs{
T_{A,B|C,D}(\l,\lb) \equiv {1\over \llb^3}T_{A,B|C,D},\quad S^m_{A,B,C|D,E}(\l,\lb) \equiv {1\over
\llb^2}S^m_{A,B,C|D,E},
}
leads to
\eqn\theortwo{
\langle T_{A,B|C,D}(\l,\lb)\rangle_{(3,2)} = \langle T_{A,B|C,D}\rangle_{(0,2)},\quad
\langle S^m_{A,B,C|D,E}(\l,\lb)\rangle_{(3,2)} = 8\langle S^m_{A,B,C|D,E}\rangle_{(0,2)}\, ,
}
where the factor $8$ comes from $\langle \ldots\rangle_{(1,2)} = 8 \langle \ldots\rangle_{(0,2)}$. As
we will see, the vector building block contributing to two-loop amplitudes with five or more particles
is
\eqn\Tmdef{
T^m_{A,B,C|D,E} \equiv A^m_A T_{B,C|D,E} + A^m_B T_{A,C|D,E} + A^m_C
T_{A,B|D,E} + 8 S^m_{A,B,C|D,E} \ ,
}
where the factor of $8$ is due to the use of \theortwo. By \jacId\ and \fivesymmE, it obeys the same
symmetry properties as $S^m_{A,B,C|D,E}(\l,\lb)$,
\eqn\symTm{
  T^m_{A,B,C|D,E} =   T^m_{(A,B,C) | (D,E)}\,,\quad
T^m_{B,D,E|A,C} =  T^m_{A,B,C|D,E}
-  T^m_{A,D,E|B,C} -   T^m_{C,D,E|A,B}\,.
}
Hence, the manipulations shown in \whyjac\ carry over to $S^m_{A,B,C|D,E}(\l,\lb) \rightarrow T^m_{A,B,C|D,E}$.

\subsec The four-point amplitude

According to the general formula \twogeneral\ the four-point amplitude at two loops is
given by
\eqn\fourtwotmp{
M_4^{(2)} =
\halfap{14}\!\!\!{\k^4 e^{2\l}\over (2\pi)^{38}2^5 192^6}  \int {d^2\Omega\over
Z_2^{22}} \int_{\Sigma_4}
|\langle D_{(14,14)} U_1 U_2 U_3 U_4\rangle_{(3,2)} |^2\Big \langle
\prod_{j=1}^4 e^{k^j\cdot x^j}\Big\rangle \, ,
}
since the four vertices cannot provide enough $d_\a$ zero-modes to 
saturate the terms with $D^m_{(13,14)}$ and $D^m_{(14,13)}$ in \caltwodef\ \NMPS.
Using the formula \DIntsOne, the Jacobi identity \jacId\ and the definitions
\theortwo\
it is straightforward to verify that (see \whyjacobi)
\eqn\fourtwokin{
\langle D_{(14,14)} U_1 U_2 U_3 U_4\rangle_{(3,2)} 
=- 96^2 c_d^2\halfap4 \big[
\langle T_{1,2|3,4}\rangle_{(0,2)} \Delta_{41}\Delta_{23}
+ \langle T_{1,4|2,3}\rangle_{(0,2)} \Delta_{12}\Delta_{34}
\big]\,.
}
Together with \Nthreezero\ and \theortwo, this implies that
\eqn\tmptwo{
M_4^{(2)} = (2\pi)^{10}\d^{10}(k)\halfap5\!{\k^4 e^{2\l}\over 2^{45}\,3^6\,5^2\pi^6}
\int {d^2\Omega \over (\det\Im\Omega)^5}
\int_{\Sigma_4}\!\! |\langle {\cal K}^{(2)}(z_1, \ldots,z_4) \rangle |^2\, \cI_4^{(2)}\, ,
}
where
\eqn\calTfour{
\langle {\cal K}^{(2)}(z_1, \ldots,z_4) \rangle \equiv \langle T_{1,2|3,4}\rangle \Delta_{41}\Delta_{23}
+ \langle T_{1,4|2,3}\rangle \Delta_{12}\Delta_{34}\,.
}
In absence of singularities $|z_{ij}|^{-2}$, using
Riemann's bilinear identity \RieOmega\ in the form of
\eqn\Riefour{
\int_{\Sigma_4} \Delta_{12} \Delta_{34} \bar \Delta_{12} \bar \Delta_{34} = 2^6 (\det \Im \Omega)^2\,,
\quad \int_{\Sigma_4} \Delta_{12} \Delta_{34} \bar \Delta_{41} \bar \Delta_{23} = 2^5 (\det \Im
\Omega)^2\,,
}
leads to the following low-energy limit
\eqn\LEfour{
\int_{\Sigma_4}| \langle {\cal K}^{(2)}(z_1, \ldots,z_4) \rangle |^2\,\cI_4^{(2)} = 2^5(\det\Im\Omega)^2\,\cK_4^{(2)} +\cO(\ap^2)\, ,
}
where (note $[\cK_4^{(2)}]= -12$)
\eqn\calFtwodef{
\cK_4^{(2)} = |\langle T_{1,2|3,4}\rangle|^2 + |\langle T_{1,4|2,3}\rangle|^2 +|\langle T_{1,3|4,2}\rangle|^2\,.
}
Finally, using the volume of the genus-two moduli space $\int d\mu_2 = 2^2\pi^3/(3^3\,5)$, one arrives
at the following low-energy limit
\eqnn\LEfourA
$$\eqalignno{
M_4^{(2)} &= (2\pi)^{10}\d^{10}(k)\halfap5\!{\k^4 e^{2\l}\over 2^{38}\,3^9\,5^3\pi^3}\,\cK_4^{(2)}
+ {\cal O}(\ap^6) \cr
 &= (2\pi)^{10}\d^{10}(k) \halfap5\!{\k^4 e^{2\l}\over 2^{10}\,3^3\,5\,\pi^3}
(s_{12}^2 + s_{13}^2 + s_{14}^2)\,\cK_4^{(1)} + {\cal O}(\ap^6) \ . &\LEfourA
}$$
In the second line we used the BRST cohomology manipulation \refs{\coefftwo,\mafraids}\foot{The normalization
of $T_{1,2|3,4}$ here is two times bigger than in \coefftwo, see definition \MDs.}
\eqn\twoloopid{
\langle T_{1,2|3,4}\rangle = 2^{14} 3^3 5\,s_{12}\langle V_1 T_{2,3,4}\rangle
}
together with the definition \forty\ of the one-loop kinematic factor ${\cal K}^{(1)}_4$ (which in turn agrees with the tree-level kinematic factor \cKfourzero).

An alternative presentation of the four-point two-loop amplitude follows by plugging the
result \twoloopid\ in \tmptwo\ and using the definition \dhokerS
\eqn\calYdef{
{\cal Y}(z_1, \ldots,z_4) \equiv s_{12} \Delta_{41}\Delta_{23}
+ s_{14} \Delta_{12}\Delta_{34}
}
to obtain
\eqn\dhoker{
M_4^{(2)} = (2\pi)^{10}\d^{10}(k)\halfap5{\k^4 e^{2\l}\over 2^{17} \pi^6}\,\cK^{(1)}_4\!\!
\int {d^2\Omega \over (\det\Im\Omega)^5}
\int_{\Sigma_4}|{\cal Y}(z_1, \ldots,z_4)|^2 \cI_4^{(2)}\,.
}
In the low-energy limit where $\cI_4^{(2)}\rightarrow 1$, using \dhokerS
\eqn\IntY{
\int_{\Sigma_4}|{\cal Y}(z_1, \ldots,z_4)|^2 = 2^5 (s_{12}^2 + s_{13}^2 + s_{14}^2)(\det\Im\Omega)^2
}
and $\int d\mu_2 = 2^2\pi^3/(3^3\,5)$ leads to the same answer \LEfourA.

\subsec The five-point amplitude

The five-point amplitude following from the general formula \twogeneral\
is given by
\eqn\amplitudetwo{
M^{(2)}_5 =  \halfap{14}\!\!\!{\kappa^5 e^{2\l}\over (2\pi)^{38}\,2^5\, 192^{6}}
\int {d^2\Omega \over Z_2^{22}}
\int_{\Sigma_5}
|\langle \langle{\cal K}^{(2)}_{[d]}(z_1, \ldots,z_5)\rangle \rangle_{(3,2)}|^2\,
\Big \langle\!\prod_{j=1}^5 e^{ k^j\cdot x^j}\Big \rangle\, ,
}
where
\eqn\caltwoFive{
\cK^{(2)}_{[d]}(z_1, \ldots,z_5) = \big(D_{(14,14)}
+ \Pi_m^1 D^m_{(13,14)} + \Pi_m^2 D^m_{(14,13)}\big)
U_1(z_1) U_2(z_2) \cdots U_5(z_5) \ .
}
For the first term in \caltwoFive, the external vertices must contribute four $d_\a(z)$ variables to
saturate the remaining $(2,2)_d$ zero-modes required by the $D_{(14,14)}$ integration. This admits one
OPE \UUOPE\ resulting in a two-particle superfield $W_{ij}^\alpha$ from \twopart\ accompanied by the singular function
$\eta_{ij}\sim z_{ij}^{-1}$ defined in \defeta. 

However, the OPE \UUOPE\ only determines the residue of
the simple pole $z_{ij}^{-1}$ and allows for two inequivalent functions of the worldsheet positions;
either $\partial_iG_{ij} \omega_I(z_j)$ or $-\partial_jG_{ij} \omega_I(z_i)$. Their difference is
regular in $z_{ij}$ and drops out from the low-energy behavior of the amplitude due to
the factor of $\delta^2(z_i-z_j)$ in \etares. Since the ambiguity does not affect the subsequent
low-energy analysis, we will use the notation $(dW_{ij})\eta_{ij}$ to leave the subtlety in the exact dependence on
$z_i,z_j$ undetermined. 

Another possible obstruction to extend the current analysis beyond the low-energy limit might stem from
OPE singularities between the $b$-ghost and the vertex operators (see for instance
\refs{\WittenCIA,\yuri}). By arguments similar to \threeloop, these might affect the two-loop
five-point amplitude at order $D^4R^5$.

Similarly, for the last two terms in \caltwoFive, the vertices must provide five $d_\a$ variables to saturate either
$(2,3)_d$ or $(3,2)_d$ zero-modes. Together with the contributions from the previous paragraph, we arrive at
\eqnn\Ubb
$$\eqalignno{
U_1U_2U_3U_4U_5\big|_{d^4}  &= \halfap5 \bigl[(dW_{12})(dW_3)(dW_4)(dW_5)\,\eta_{12} + (1,2|1,2,3,4,5)\bigr]  &\Ubb\cr
 & + \halfap4 \sum_{I=1}^2\Pi^m_I \omega_I(z_1) A^1_m (dW^2)(dW^3)(dW^4)(dW^5) + (1\leftrightarrow
 2,3,4,5)\,,\cr
U_1U_2U_3U_4U_5\big|_{d^5} &= \halfap5 (dW_1)(dW_2)(dW_3)(dW_4)(dW_5) \ .
}$$
Using the formulas in \DIntsOne\ a long but straightforward calculation leads to
\eqn\Dff{
\langle \langle \cK^{(2)}_{[d]}(z_1, \ldots,z_5)\rangle \rangle_{(3,2)}
= \halfap4 96^2 c_d^2\,\langle\cK^{(2)}(z_1, \ldots,z_5)\rangle_{(0,2)}\, ,
}
where
\eqnn\Ldelta
$$\eqalignno{
{\cal K}^{(2)}(z_1, \ldots,z_5) 
& \equiv  \big[ T^m_{1,2,3|4,5} \sum_{I=1}^2 \Delta_{51}\omega_I(z_2)\Delta_{34}\Pi_{m}^I
+ {\rm cyc(12345)}\big]
 &\Ldelta
\cr
+\halfap{}\Big[\eta_{12}& (  T_{12,3|4,5} \Delta_{24}\Delta_{35} +
 T_{12,4|3,5} \Delta_{23}\Delta_{45})  + (1,2|1,2,3,4,5)\Big]\,,
}$$
and $T^m_{1,2,3|4,5}$ is defined in \Tmdef. As detailed in \whyjac, the symmetry property \symTm\ of
$T^m_{1,2,3|4,5}$ is crucial to obtain the first line of \Ldelta\ in terms of two factors of
$\Delta_{ij}$. After discarding $\Pi_m^I \rightarrow 0$, \Ldelta\ is the low-energy regime of the open-string
worldsheet integrand for the five-point two-loop amplitude.

Collecting the above results, the low-energy regime of the two-loop amplitude reads
\eqn\Asix{\eqalign{
M_5^{(2)} &=
\halfap6\! {\kappa^5 e^{2\l}\over 2^{37} \, 3^{2}\,\pi^6}
\int {d^2\Omega \over (\det\Im\Omega)^5}\int_{\Sigma_5}
\bigl|\langle \cK^{(2)}(z_1, \ldots,z_5) \rangle_{(0,2)}\bigr|^2 \Big \langle \KN5\Big \rangle
+ {\cal O}(\alpha'^7)
\cr
&=
(2\pi)^{10}\d^{10}(k)\halfap5\! {\kappa^5 e^{2\l}\over 2^{45} \, 3^{6}\,5^2\,\pi^6}
\int {d^2\Omega \over (\det\Im\Omega)^5}\int_{\Sigma_5}
\bigl|\langle \cK^{(2)}(z_1, \ldots,z_5)  \rangle\bigr|^2 \cI_5^{(2)} + {\cal O}(\alpha'^7) \ ,
}}
where in the second line we used \Nthreezero.

\subsubsec The low-energy limit

The low-energy limit of the genus-two integral in \Asix\ can be extracted along the same
lines as done at genus one. First of all, \LRcontract\ allows to perform contractions
among left- and right-moving zero-modes of $\Pi^m$ which can be integrated over $\Sigma_5$ using
\eqnn\fivecycA
$$\eqalignno{
\int_{\Sigma_5}\Delta_{12}\sum_{I=1}^2 \Pi^I_m \omega_I(z_3)\Delta_{45}\times
\bar \Delta_{12}\sum_{J=1}^2\bar \Pi^J_n \bar \omega_J(\bar z_3)\bar\Delta_{45}&= -2^8\pi \, \halfap \,  \eta_{mn}(\det\Im\Omega)^2 \cr
\int_{\Sigma_5}\Delta_{12}\sum_{I=1}^2 \Pi^I_m \omega_I(z_3)\Delta_{45}\times
\bar \Delta_{34}\sum_{J=1}^2\bar \Pi^J_n \bar \omega_J(\bar z_5)\bar\Delta_{12}&=2^7 \pi \, \halfap \,  \eta_{mn}(\det\Im\Omega)^2 &\fivecycA \cr
\int_{\Sigma_5}\Delta_{12}\sum_{I=1}^2 \Pi^I_m \omega_I(z_3)\Delta_{45}\times
\bar \Delta_{23}\sum_{J=1}^2\bar \Pi^J_n \bar \omega_J(\bar z_4)\bar\Delta_{51}&= -2^6\pi \, \halfap \,  \eta_{mn}(\det\Im\Omega)^2 
}$$
and cyclic permutations. Then, the subset of the terms $\sim \eta_{ij} \bar \eta_{pq}$ in \Asix\ with
``diagonal'' labels $i=p$ and $j=q$ contributes according to \etares, resulting in ten permutations of
$$
\int_{\Sigma_4} \big|
\langle T_{12,3|4,5}\rangle \Delta_{24}\Delta_{35} +
\langle T_{12,4|3,5}\rangle \Delta_{23}\Delta_{45}
\big|^2 = 2^5 (\det \Im \Omega)^2 \big[ | \langle T_{12,3|4,5} \rangle |^2 + {\rm cyc}(3,4,5) \big]\, ,
$$
where the integrals are identical to the four-point case \LEfour. Permutations of $\eta_{12} \bar \eta_{13}$ or
$\eta_{12} \bar \eta_{34}$ from the holomorphic square in \Ldelta\ do not contribute to the low-energy
limit \Asix.

By assembling the two sectors with and without contractions between left- and right-movers,
one can show that the leading-order terms of the five-point two-loop amplitude \Asix\ are given
by
\eqn\Les{
\int_{\Sigma_5} |\langle \cK^{(2)}(z_1, \ldots,z_5)\rangle|^2 {\cal I}_5^{(2)}= 2^6\pi\halfap{}(\det\Im\Omega)^2\, \cK_5^{(2)} + {\cal O}(\ap^2)\,,
}
with kinematic factor (note $[\cK_5^{(2)}]=-12$)
\eqn\calKFiveTwo{
\cK_5^{(2)} \equiv
{\big|\langle T_{12,3|4,5}\rangle \big|^2\over s_{12}} + {\big|\langle T_{12,4|3,5}\rangle \big|^2\over s_{12}}
+ {\big|\langle T_{12,5|3,4}\rangle \big|^2\over s_{12}} +
\big|\langle T^m_{3,4,5|1,2}\rangle\big|^2 +
(1,2|1,2,3,4,5) \ .
}
Hence, using  $\int d\mu_2 =
2^2\pi^3/(3^3\,5)$ implies the following low-energy limit of \Asix,
\eqnn\LEtmp
$$\eqalignno{
M_5^{(2)} &= (2\pi)^{10}\d^{10}(k)\halfap6\!{\kappa^5 e^{2\l} \over 2^{37} \, 3^{9}\,5^3\,\pi^2}\,
\cK_5^{(2)} + \cO(\ap^7) \ . &\LEtmp
}$$

\subsubsec Components in type IIB and type IIA

A long and tedious calculation \PSS\ identifies the type IIB components of the two-loop
kinematic factor \calKFiveTwo\ with the $\alpha'$-correction $\sim \zeta_5$ of the
five-point tree-level amplitude \fivetree\ and \KLT:
\eqn\KLTfive{
\cK_5^{(2)} \big|_{{\rm IIB}}= -2^{28}\,3^6\,5^2 \halfap{-5} \cK_5^{(0)}\Big|_{\zeta_5}
 \times \cases{\ \ \, 1 \,  \ : \ {\rm five} \  {\rm gravitons} \cr -{3\over 5} \ : \ {\rm four} \ {\rm gravitons}, \ {\rm one} \ {\rm dilaton}}
}
Similar to the kinematic factor \CompFive\ in the one-loop low-energy limit, the relative
coefficient to the tree-amplitude depends on the total R-symmetry charge of the external states,
in lines with S-duality. The components considered in \KLTfive\ extend to a variety of
further state combinations of alike R-symmetry charges by linearized supersymmetry.

Similar to the analogous one-loop result \CompFiveA, type IIA components involve additional tensor
structures as compared to the $A^{\rm YM}$ bilinears in the tree-amplitude,
\eqn\typeIIA{
\cK_5^{(2)} \big|^{5 \ {\rm gravitons}}_{{\rm IIA}}= -2^{28}\,3^6\,5^2 \halfap{-5}\Big[ \cK_5^{(0)}\Big|_{\zeta_5}
-  {1\over 2} \sum_{1\leq i < j}^5 s_{ij}^2\big|\epsilon^{me^1k^2e^2k^3e^3k^4e^4k^5e^5}\big|^2 \Big] \ ,
}
where $\epsilon^{me^1k^2e^2k^3e^3k^4e^4k^5e^5}\equiv
\epsilon_{10}^{mn p_2q_2\ldots p_5 q_5} e_n^1 k_{p_2}^2 e_{q_2}^2\ldots k_{p_5}^5 e_{q_5}^5$.

Upon insertion into \LEtmp, the kinematic factors \KLTfive\ and \typeIIA\ give rise to the following
low-energy limits for the five-graviton amplitudes:
\eqnn\LEtmpfinalB
$$\eqalignno{
M_5^{(2)} \big|^{\ap^6}_{{\rm IIB} \ {\rm gravitons}}
&= (2\pi)^{10}\d^{10}(k)\halfap{}  {\kappa^5 e^{2\l} \over 2^{9} \, 3^{3}\,5\,\pi^2} 
 \cK_5^{(0)}\Big|_{\zeta_5}
 &\LEtmpfinalB
 \cr
 M_5^{(2)} \big|^{\ap^6}_{{\rm IIA}\ {\rm gravitons}}
&= (2\pi)^{10}\d^{10}(k)\halfap{}  {\kappa^5 e^{2\l} \over 2^{9} \, 3^{3}\,5\,\pi^2} 
\Big[ \cK_5^{(0)}\Big|_{\zeta_5}
-  {1\over 2} \sum_{1\leq i < j}^5 s_{ij}^2\big|\epsilon^{me^1k^2e^2k^3e^3k^4e^4k^5e^5}\big|^2 \Big]
}$$
According to \KLTfive, the R-symmetry violating type IIB components (e.g. four gravitons and one dilaton) carry an extra factor of $-{3\over5}$.

\newsec{S-duality properties}
\par\seclab\secSdual

\noindent In this section we are going to show that the type IIB five-point amplitudes computed with the non-minimal
pure spinor formalism agree with expectations based on S-duality.

\subsec Review of four-point S-duality

In the string frame, the $SL(2,\Bbb Z)$-duality prediction for the perturbative four-graviton
type IIB effective action is given by \refs{\GGRq,\GreenKVan,\Greenthreeloop}
\eqnn\Action
$$\eqalignno{
S_{\rm IIB}^{4{\rm pt}} = \int d^{10}x\sqrt{-g}\,\big[& R^4(2\zeta_3 e^{-2\phi} + 4 \zeta_2)
+ D^4R^4(2\zeta_5 e^{-2\phi} + {8\over 3}\zeta_4e^{2\phi}) &\Action\cr
&+ D^6R^4(4\zeta_3^2 e^{-2\phi} + 8\zeta_2\zeta_3 + {48\over 5}\zeta_2^2 e^{2\phi} + {8\over
9}\zeta_6 e^{4\phi}) + \cdots \big],
}$$
where the ellipsis refers to terms of higher order $D^{\geq 8}R^4$.
A dilaton dependence of the form $e^{(2g-2)\phi}$ is associated with the $g$-loop 
order in string perturbation theory. The tensor structure of the covariant derivatives $D$ 
and Riemann curvature tensors $R$ suppressed in the shorthands $R^4$, $D^4R^4$ and
$D^6R^4$ will not be important in the following. The coefficients of the $R^4$
and $D^4R^4$ interactions can be identified with the zero-modes of the non-holomorphic Eisenstein series
\eqnn\eisth
\eqnn\eisfh
$$\eqalignno{
E_{3/2}(\Phi,\bar \Phi) &\equiv 2\zeta_3 e^{-3\phi/2} + 4 \zeta_2 e^{\phi/2} + \cdots
&\eisth \cr
E_{5/2}(\Phi,\bar \Phi) &\equiv 2\zeta_5 e^{-5\phi/2} + {8\over 3}\zeta_4e^{3\phi/2} + \cdots
&\eisfh
}
$$
depending on the complex axio-dilaton field $\Phi \equiv C_0+ie^{-\phi}$. A relative factor of
$e^{\pm\phi/2}$ stems from the transformation between string frame and Einstein frame. The Fourier
modes in the ellipsis of \eisth\ and \eisfh\ describe the non-perturbative completion of the type IIB
action \refs{\GGRq,\GreenKVan} and ensure modular invariance w.r.t. $\Phi$. The prefactor of the
$D^6R^4$ operator in \Action\ was firstly predicted in \Greenthreeloop\ and descends from a
modular-invariant function which is made explicit in \GreenYXA.

The four-point amplitudes reviewed in the previous sections exhibit the following low-energy behavior
(in both type IIB and type IIA theory):
\eqnn\FourAmps
$$\eqalignno{
M^{(0)}_4 &= (2\pi)^{10}\d^{10}(k)\halfap3 \kappa^4 e^{-2\l}\,2\pi \cK_4^{(0)}\big(
{3\over \s_3} + 2\zeta_3 + \zeta_5 \s_2 + {2\over 3}\zeta_3^2 \s_3 + \cdots \big) \cr
M^{(1)}_4 &= (2\pi)^{10}\d^{10}(k)\halfap3 {\k^4 \over 2^4\,3 \pi}
\cK_4^{(0)}\big(1 + {\zeta_3\over 3}\sigma_3 + \cdots \big) &\FourAmps \cr
M_4^{(2)} &= (2\pi)^{10}\d^{10}(k)\halfap3{\k^4 e^{2\l}\over 2^{10}\,3^3\,5\pi^3}\,
\cK^{(0)}_4 \big(\sigma_2 +\cdots \big) \cr
%
}$$
The loop- and $\alpha'$-orders are in one-to-one correspondence with the curvature
couplings $e^{(2g-2)\phi} D^{2k} R^4$ in the action \Action. Matching the ratio
of the $R^4$ interactions $\sim \zeta_3$ and $\sim \zeta_2$ with the values computed
in \FourAmps\ relates the coupling constants $e^\phi$ and $e^\l$,
\eqn\Rfourratio{
{e^{2\phi}\pi^2\over 3\zeta_3} = {e^{2\l}\over 2^6\,3\pi^2\zeta_3} \rightarrow e^{2\l} = 2^6 \pi^4\,e^{2\phi} \ .
}
Furthermore, one can verify using the conversion factor \Rfourratio\ that the ratio of all the
interactions match between their predicted values in the action \Action\ and the explicit amplitude
computations summarized in \FourAmps. The first perturbative verification of the expressions in
\Action\ was achieved in \refs{\GGRq,\GreenOneLoopcoeff} for genus one, in
\refs{\dhokerVI,\dhokerS,\DHokerEEA} for genus two and \threeloop\ for genus three.

\subsec S-duality at five-points for graviton couplings

We will now check if the above ratios predicted for the four-point amplitudes at different loop orders
also hold for their corresponding five-point amplitudes at one- and two-loops. The extension of the
type IIB effective action \Action\ beyond the four-point level complements the four-curvature
corrections $D^{2k} R^{4}$ by a tail of operators\foot{In addition, novel couplings of the form
$D^{2k}R^{\geq 5}$ without a four-field representative in their supersymmetric completion might
arise, e.g. the $D^6R^5$ interaction identified at one-loop \FiveSdual.} $D^{2(k-l)} R^{4+l}$ with higher powers of curvature $l=1,2,\ldots,k$ required by non-linear supersymmetry. 

The result for the two-loop five-point amplitude confirms that the five-field completion $(D^{4} R^{4}+D^{2} R^{5})$ is accompanied uniformly by the zero-modes of $E_{5/2}$ given in \eisfh.
Similarly, the compatibility of the $E_{3/2}R^4$ interaction with five-point amplitudes was
verified through the one-loop analysis in \FiveSdual. These checks are based on the $\ap$-expansion of
the five-point IIB amplitudes at tree-level, one- and two-loop computed in the previous sections,
\eqnn\FiveAmps
$$\eqalignno{
M^{(0)}_5 &= (2\pi)^{10}\d^{10}(k)\halfap{}\; \kappa^5 e^{-2\l}(2\pi)^2 \cK_5^{(0)} &\FiveAmps\cr
M^{(1)}_5 \big|^{\ap^4}_{{\rm IIB}}&= (2\pi)^{10}\d^{10}(k)\halfap{} {\k^5 \over  2^{4} 3 }\, \cK_5^{(0)}\Big|_{\zeta_3} \times \cases{\ \ \, 1 \,  \ : \ {\rm five} \  {\rm gravitons} \cr -{1\over3} \ : \ {\rm four} \ {\rm gravitons}, \ {\rm one} \ {\rm dilaton}}\cr
M^{(2)}_5 \big|^{\ap^6}_{{\rm IIB}}&= (2\pi)^{10}\d^{10}(k)\halfap{}  {\kappa^5 e^{2\l} \over 2^{9} \, 3^{3}\,5\,\pi^2} \cK_5^{(0)}\Big|_{\zeta_5}\times \cases{\ \ \, 1 \,  \ : \ {\rm five} \  {\rm gravitons} \cr -{3\over 5} \ : \ {\rm four} \ {\rm gravitons}, \ {\rm one} \ {\rm dilaton}} \ ,
}$$
where the tree-level factor $\cK_5^{(0)}$ is given by \KLT\ \motivic.
Hence, the ratios of the corresponding five-point interactions at one-loop
are easily checked to agree with the perturbative terms in the Eisenstein series \eisth\ and \eisfh,
\eqn\fiveRO{
{M^{(1)}_5\over M^{(0)}_5} \Big|^{\ap^4}_{{\rm IIB} \ {\rm gravitons}}= {e^{2\l}\over 2^6\,3\,\pi^2\zeta_3} = {2e^{2\phi}\zeta_2\over \zeta_3} \ ,
}
and similarly at two-loops (recall that $\zeta_2 = {\pi^2\over6}$ and $\zeta_4 = {\pi^4\over90}$),
\eqn\fiveRT{
{M^{(2)}_5\over M^{(0)}_5}\Big|^{\ap^6}_{{\rm IIB} \ {\rm gravitons}} = {e^{4\l}\over 2^{11}\,3^3\,5\pi^4\zeta_5} =
{4e^{4\phi}\zeta_4\over 3\zeta_5}\,.
}
By modular invariance of the Eisenstein series, \fiveRO\ and \fiveRT\ confirm S-duality at the five-point level.

\subsec S-duality at five-points for dilaton couplings
\par\subseclab\secDilatonS

\noindent The ratios of tree-level and loop-amplitudes seen in \FiveAmps\ depend on the external type IIB states,
i.e. trading one of the five gravitons for a dilaton introduces additional factors of $-{1\over 3}$ and
$-{3\over 5}$ into the comparison of low-energy limits. These numbers have a natural explanation from
the Einstein frame presentation of the leading terms in \Action,
\eqn\Einsteinframe{
R^4E_{3/2}(\Phi,\bar \Phi) + (D^4R^4 + D^2 R^5)E_{5/2}(\Phi,\bar \Phi) +\cdots \, ,
}
see \eisth\ and \eisfh\ for their perturbative contributions.

Processes which violate the R-symmetry of type IIB supergravity (such as the scattering of four gravitons
and one dilaton) are associated with operators which transform with modular weight under S-duality
\Rviolating. Hence,
by modular invariance of the type IIB action, they must accompanied by modular forms of opposite weights.
The latter can be obtained from modular invariant functions such as $E_s$ by acting with the modular covariant derivative
\eqn\modcov{
{\cal D}: \ e^{q\phi} \rightarrow q \cdot e^{q\phi} \ .
}
The modular forms obtained from $E_{3/2}$ and $E_{5/2}$ are characterized by the following perturbative terms (with Fourier-modes in the ellipsis):
\eqnn\Rvio
\eqnn\Rvioa
$$\eqalignno{
{\cal D}E_{3/2}(\Phi,\bar \Phi) &=  \Big(-{3\over 2} \Big)2\zeta_3 e^{-3\phi/2} +\Big({1\over 2} \Big)
4 \zeta_2 e^{\phi/2}+\cdots
&\Rvio
\cr
{\cal D} E_{5/2}(\Phi,\bar \Phi) &=
\Big(-{5\over 2} \Big)2\zeta_5 e^{-5\phi/2} +\Big({3\over 2} \Big) {8\over 3}\zeta_4e^{3\phi/2} +\cdots \ .
&\Rvioa
}$$
In comparison with \eisth\ and \eisfh, the ratio between tree-level and higher-genus contributions is
deformed by the covariant derivative in \modcov, namely by $-{1\over 3}$ and $-{3\over 5}$ in cases of
$E_{3/2}$ and $E_{5/2}$, respectively. The modular forms in \Rvio\ and \Rvioa\ multiply the
R-symmetry violating counterparts of $R^4$ and $(D^4R^4 + D^2 R^5)$ interactions
which in turn describe the dilatonic amplitude components in \FiveAmps. Hence, the covariant derivative
in \modcov\ holds the key for the S-duality origin of the relative factors between graviton and dilaton
amplitudes in \FiveAmps. It would be interesting to extend the analysis to higher orders in $\alpha'$
and to compare the ratios between amplitudes at tree-level and two-loops for higher derivative
operators with and without R-symmetry charges, as it was done in \FiveSdual\ at one-loop up to order
$(\ap)^9$.


\newsec{Conclusion}

As the main result of this work, we have computed the low-energy limit of the five-point two-loop
amplitude among massless type II closed-string states. The superspace representation of the result is
given in \calKFiveTwo\ with prefactors made precise in \LEtmp. The type IIB components involving five gravitons as well as
four gravitons and one dilaton were found to match the tree-level amplitude at the corresponding
order in $\ap$, see \KLTfive. The determined ratios tie in with the
S-duality expectation based on the $E_{5/2}$ coefficient of the $(D^4 R^4+D^2R^5)$ operator in the
effective action \GreenKVan\ and its counterpart ${\cal D} E_{5/2}$ with modular weight, see \Rvioa.

The computation was performed using the non-minimal pure spinor formalism \NMPS\ where the
normalizations can be reliably kept track of and where the $b$-ghost is explicitly known. However,
subtle issues regarding possible OPE singularities between the $b$-ghost and the vertex operators (see
for instance \WittenCIA) currently prevent the determination of the five-point two-loop amplitude to
all orders in $\ap$. These subtleties did not affect the two-loop low-energy analysis of this work, but
it would certainly be desirable to extend the five-point correlator in \Ldelta\ to all orders in the
low-energy expansion. Starting from the Zhang-Kawazumi invariant expected at the subleading order in
$\ap$ \DHokerEEA, the systematics of the low-energy expansion and the threshold corrections deserve to
be studied along the lines of the one-loop results in \GreenOneLoopcoeff. The $\alpha'$-expansion of
the corresponding open string amplitudes at two-loops calls for a higher-genus generalization of the
elliptic multiple zeta values \Enriquez\ which were studied in the context of planar one-loop
amplitudes in \BroedelVLA.

Also, it would be rewarding to cast the kinematic factors into the language of the minimal pure spinor
superspace of \superpoincare\ and to bypass the computational steps required by the extra worldsheet
variables of the non-minimal pure spinor formalism. In particular, this concerns the evaluation of covariant
derivatives originating from $r_\alpha$ and the tensor manipulations required to arrange the $\bar
\lambda_\alpha$ into contractions with $\lambda^\alpha$.
For the three-loop four-point kinematic factors of \threeloop, a much simpler
BRST-equivalent representation in terms of (minimal) pure spinor superspace has recently
been found \HighSYM.

\bigskip \noindent{\bf Acknowledgements:} We thank Nathan Berkovits, Pierre Vanhove and especially
Michael Green for valuable discussions. We also thank Arnab Rudra for useful comments on the draft. CRM and OS 
acknowledge support by the European Research Council Advanced Grant No. 247252 of Michael Green. CRM
thanks AEI and the Perimeter Institute for hospitality, OS is grateful to DAMTP for kind hospitality during finalization of this work, and HG is grateful to the Perimeter Institute for warm hospitality and partial financial support during
stages of this work. HG is supported by FAPESP grant 2011/13013-8.

\appendix{A}{Worldsheet factorization of the amplitudes}
\applab\appUNI

\noindent In this appendix we show that five-point multiloop amplitudes computed in main body of this
work factorize correctly on their massless poles as required by unitarity. We first fix the overall
normalization $\kappa$ of the vertex operators by imposing unitarity for the four-point amplitude at
tree-level. After that there is no freedom left to adjust parameters and we proceed to check the
factorization of the higher-loop amplitudes.

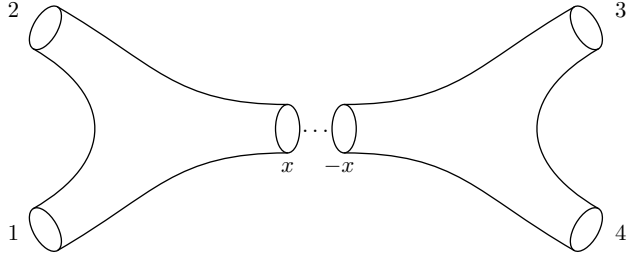
\ifig\figfourfac{The factorization of the tree-level four-point amplitude in the
massless pole $s_{12}$ in terms of three-point amplitudes. This condition was used to fix the normalization
constant $\kappa^2 = {\pi e^{2\l}\over \ap^2}$ in equation \kappaUn.}
{\epsfxsize=0.50\hsize\epsfbox{5pt_factorization.4}}

\subsec Factorization of the four-point tree-level amplitude

The factorization constraint for the massless pole $s_{12}$ in the four-point amplitude
reads
\eqn\UniCon{
M^{(0)}_4\Big|_{s_{12}} = \ap^4 \int {d^{10}k\over (2\pi)^{10}} \sum_x {M^{(0)}_3(1,2,x) M^{(0)}_3(-x,3,4)\over k_x^2}
}
where the notation $\big|_{s_{12}}$ projects to the pole in $s_{12}$ and
discards regular terms in $s_{12}$, the sum $\sum_x$ runs
over all states $x$ in the supergravity multiplet and $-x$ represents the state~$x$ at
momentum $-k_x$ and complex conjugate polarization. This is depicted in \figfourfac.

On the one hand, recall the low-energy limit \fourlow\ of the four-point amplitude
\eqn\fourpoleAmp{
M^{(0)}_4\Big|_{s_{12}} = (2\pi)^{10}\d^{10}(k)\k^4 e^{-2\l}\, 2\pi {|\langle V_{12}V_3V_4\rangle |^2\over s_{12}} 
\,.
}
On the other hand, a short computation using the factorization constraint \UniCon\ yields
\eqnn\unitarity
$$\eqalignno{
M^{(0)}_4\Big|_{s_{12}} &= \ap^4 \int {d^{10}k\over (2\pi)^{10}} \sum_x {M^{(0)}_3(1,2,x) M^{(0)}_3(-x,3,4)\over k_x^2}\cr
&= (2\pi)^{10}\d^{10}(k)\k^6 e^{-4\l}\halfap{-2} \ap^4 {1\over 2s_{12}}|\langle V_{12}V_3V_4\rangle |^2\,,
&\unitarity
}$$
where the three-point amplitude is given by
\eqn\againthree{
M^{(0)}_3(1,2,x) =(2\pi)^{10}\d^{10}(k^1+k^2+k^x) \halfap{-1} \kappa^3 e^{-2\l}\;|\langle V_1 V_2
V_x\rangle|^2 \ .
}
and we used
\eqn\UnitaryInt{
\int d^{10}k {\d^{10}(k^1+k^2+k^x)\d^{10}(-k^x + k^3+\cdots +k^n)\over k^2_x} = {1\over
2s_{12}}\d^{10}(k^1+k^2+k^3+\cdots +k^n)
}
together with the explicit component sum \PSS
\eqn\treefact{
\sum_x \langle V_1V_2V_x\rangle\langle V_x V_3V_4\rangle = \langle V_{12}V_3V_4\rangle  + {\cal
O}(s_{12})\,.
}
Therefore equating \fourpoleAmp\ and \unitarity\ leads to
\eqn\kappaUn{
\kappa^2 e^{-2\l} = {\pi\over \ap^2}\,.
}

\subsec Factorization of the five-point tree-level amplitude

The normalization of the five-point tree amplitude \fivetree\ will be checked through
its factorization on the massless $s_{12}$ pole according to
\eqn\fiveUn{
M^{(0)}_5\Big|_{s_{12}} = \ap^4 \int {d^{10}k\over (2\pi)^{10}}  \sum_x { M^{(0)}_3(1,2,x) M^{(0)}_4(-x,3,4,5)  \over
k^2_x} \ ,
}
where the three-point amplitude was recalled in \againthree\ and 
\eqnn\instead
$$\eqalignno{
M^{(0)}_4(-x,3,4,5) &= (2\pi)^{10}\d^{10}(-k^x + k^3 + k^4 + k^5)\,\k^4 e^{-2\l} &\instead\cr
&\hskip15pt \times 2\pi \Big[
{|\langle V_xV_3  V_{45}\rangle |^2\over s_{45}} + {|\langle V_xV_4V_{35}\rangle |^2\over s_{35}}
+ {|\langle V_xV_5V_{34}\rangle |^2\over s_{34}} \Big] + \cO(\ap^3) \ .
}$$
Using \kappaUn\ and
\eqn\fivefac{
\sum_x \langle V_1V_2V_x\rangle\langle V_x V_3V_{45}\rangle = \langle V_{12}V_3V_{45}\rangle + {\cal
O}(s_{12},s_{45})
}
the factorization constraint \fiveUn\ gives
\eqnn\bcjdouble
$$\eqalignno{
M^{(0)}_5\Big|_{s_{12}} ={} &(2  \pi)^{10}\d^{10}(k)\halfap{}\; \kappa^5 e^{-2\l}(2\pi)^2 &\bcjdouble\cr
& \times\Big[
  {|\langle V_{12}V_3V_{45}\rangle|^2\over s_{12}s_{45}}
+ {|\langle V_{12}V_{35}V_4\rangle|^2\over s_{12}s_{35}}
+ {|\langle V_{12}V_5V_{34}\rangle|^2\over s_{12}s_{34}}
\Big] \big|_{s_{12}} + \cO(\ap^3)\,.
}$$
This ties in with a component comparison of the terms with a pole in $s_{12}$,
\eqn\compInd{
\tilde A^T_{54}\cdot S_0\cdot  A_{45} \big|_{s_{12}} =
\Big[  {|\langle V_{12}V_3V_{45}\rangle|^2\over s_{12}s_{45}}
+ {|\langle V_{12}V_{35}V_4\rangle|^2\over s_{12}s_{35}}
+ {|\langle V_{12}V_5V_{34}\rangle|^2\over s_{12}s_{34}}
\Big]\big|_{s_{12}} \ ,
}
which confirms the normalization of the five-point closed-string amplitude \fivetree.

\subsec Factorization of the five-point one-loop amplitude

From the low-energy limit of the five-point amplitude \LowEoneFinal, it follows that
\eqnn\LowFiveAp
$$\eqalignno{
M^{(1)}_5\Big|_{s_{12}} &=
(2\pi)^{10}\d^{10}(k)\halfap{4} {\k^5 \over  2^{9}\, 5^2\,3}\, {|\langle T_{12|3,4,5}\rangle|^2\over s_{12}}
 + \cO(\ap^5)\,,\cr
& =
(2\pi)^{10}\d^{10}(k)\halfap{4} {\k^5 \over  2^{3}\, 3} {|\langle V_{12}T_{3,4,5}\rangle |^2\over
s_{12}} + {\cal O}(\ap^5)&\LowFiveAp\,,
}$$
where in the second line we
used $\langle T_{12|3,4,5}\rangle = 40 \big[ \langle V_{12}T_{3,4,5}\rangle  - {s_{12}\over 11}\langle A_{12|3,4,5}\rangle\big]$
(as shown in the appendix~\appBRST) and discarded the contact term since it does
not contribute to the $s_{12}$ pole.

One the other hand, given the three-point tree \againthree\ and the low-energy limit of \fourtmp,
\eqnn\LEs
$$\eqalignno{
M_4^{(1)}  &= (2\pi)^{10}\d^{10}(k){\kappa^4\over 2^4\,3\,\pi}\halfap3 |\langle V_1 T_{2,3,4}\rangle|^2  + {\cal O}(\ap^4) \ , &\LEs
}$$
the factorization constraint
\eqn\UnitFive{
 M_5^{(1)}\Big|_{s_{12}} = \ap^4 \int {d^{10}k\over (2\pi)^{10}} \sum_x {M_3^{(0)}(1,2,x)M^{(1)}_4(-x,3,4,5)\over k^2_x}
}
together with \UnitaryInt\ yields
\eqn\factsol{
 M_5^{(1)}\Big|_{s_{12}} = (2\pi)^{10}\d^{10}(k) \ap^4 \halfap2 {\k^7 e^{-2\l} \over 2^5\,3\,\pi}
{1\over s_{12}}
\sum_x |\langle V_1 V_2 V_x\rangle|^2 |\langle V_x T_{3,4,5}\rangle|^2 + {\cal O}(\ap^7) \ .
}
One can show via a component expansion that the kinematic factors satisfies \PSS
\eqn\loopfact{
\sum_x \langle V_1 V_2 V_x\rangle \langle V_x T_{3,4,5}\rangle =
\langle V_{12}T_{3,4,5}\rangle + \cO(s_{12})\,.
}
By \kappaUn, one finally arrives at
\eqn\factsolTwo{
M_5^{(1)}\Big|_{s_{12}} = (2\pi)^{10}\d^{10}(k)\halfap4 {\k^5\over 2^3\,3}{|\langle
V_{12}T_{3,4,5}\rangle|^2 \over s_{12}}  + {\cal O}(\ap^5)\ ,
}
in complete agreement with the expression \LowFiveAp.

\subsec Factorization of the five-point two-loop amplitude

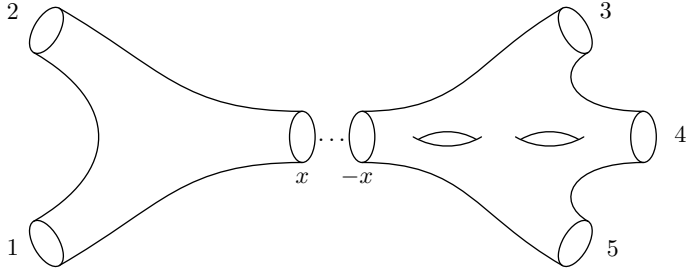
\ifig\figone{Factorization channel of the five-point two-loop amplitude into a tree-level three-point
and a four-point two-loop amplitude.}
{\epsfxsize=0.55\hsize\epsfbox{5pt_factorization.3}}

\noindent In the low-energy limit of the five-point two-loop amplitude \LEtmp, the terms with a
pole in $s_{12}$ are given by
\eqn\lowfive{
M_5^{(2)}\Big|_{s_{12}} =(2\pi)^{10}\d^{10}(k) \halfap6 {\kappa^5 e^{2\l}\over 2^{37} \, 3^{9}\,5^3\,\pi^2}\,\Big[
{|\langle T_{12,3|4,5}\rangle|^2\over s_{12}} + {|\langle T_{12,4|3,5}\rangle|^2\over s_{12}}
+ {|\langle T_{12,5|3,4}\rangle|^2\over s_{12}}\Big]\,.
}
The $s_{12}$-channel factorization constraint in the low-energy limit
\eqn\UnitFiveTwo{
 M_5^{(2)}\Big|_{s_{12}} = \ap^4 \int {d^{10}k\over (2\pi)^{10}}\sum_x {M_3^{(0)}(1,2,x)M^{(2)}_4(-x,3,4,5)\over k^2_x}
}
for the factorization
into a tree-level three-point amplitude \threeptAmp\ and a
two-loop four-point amplitude \LEfourA\
\eqnn\inputFive
$$\eqalignno{
M^{(0)}_3 &=(2\pi)^{10}\d^{10}(k) \halfap{-1} \kappa^3 e^{-2\l}\;|\langle V_1 V_2 V_3\rangle|^2
&\inputFive\cr
M_4^{(2)} &= (2\pi)^{10}\d^{10}(k)\halfap5{\k^4 e^{2\l}\over 2^{38}\,3^9\,5^3\pi^3}
\big[
|\langle T_{1,2|3,4}\rangle|^2 + |\langle T_{1,4|2,3}\rangle|^2 +|\langle T_{1,3|4,2}\rangle|^2
\big] + {\cal O}(\ap^6) \,,
}$$
yields
\eqn\inter{
 M_5^{(2)}\Big|_{s_{12}} = (2\pi)^{10}\d^{10}(k)\halfap4 { \ap^4\k^7\over
 2^{39}\,3^9\,5^3\pi^3}
\Big[
{|\langle T_{12,3|4,5}\rangle|^2\over s_{12}} + {|\langle T_{12,4|3,5}\rangle|^2\over s_{12}}
+ {|\langle T_{12,5|3,4}\rangle|^2\over s_{12}}
\big]
}
where we used \UnitaryInt\ and \PSS
\eqn\truetwo{
\sum_x \langle V_1 V_2 V_x\rangle\langle T_{x,3|4,5}\rangle = \langle T_{12,3|4,5}\rangle + \cO(s_{12})
}
Therefore the constraint \kappaUn\ ($\ap^2\k^2 = \pi e^{2\l}$) implies the agreement of \inter\ with \lowfive\ and establishes
the correct factorization of the five-point two-loop amplitude.

\appendix{B}{One-loop kinematic factors: NMPS versus MPS representation}
\applab\appBRST

\noindent From equations (3.7) and (4.7) of \oneloopNMPS\ and using the definitions \Tijkdef\ it follows
that
\eqnn\TodNMPS
$$\eqalignno{
T_{12|3,4,5} &= \llb V_{12} [ 36 \Tijk3,4,5 + 4 (\l\g^m W_4)(\l\g^n W_5)F^3_{mn}] + s_{12} J_{12|3,4,5}&\TodNMPS\cr
T_{1|23,4,5} &= \llb V_1 \left[ 36  \Tijk23,4,5 + 4 (\l\g^m W_{23})(\l\g^n W_4)F^5_{mn}\right] + s_{23}( J_{13|2,4,5} - J_{12|3,4,5})\cr
J_{12|3,4,5} &\equiv  (\l\g^m W_4)(\l\g^n W_5) \big[ V_1V_2 (\lb\g_{mn}W_3) + 2\llb V_2 (A_1\g_{mn}W_3) - (1\leftrightarrow 2)\big] \ .
}$$
Since \TodNMPSdef\ is totally symmetric in $(345)$ it is possible to rewrite \TodNMPS\ more conveniently by averaging it over its permutations
and using the Theorem~1 of \threeloop\ to factor out $\llb$,
\eqnn\TodEquiv
$$\eqalignno{
T_{12|3,4,5} &= 40\llb \big[V_{12} \Tijk3,4,5  - {1\over 11} s_{12}A_{12|3,4,5}\big]&\TodEquiv\cr
T_{1|23,4,5} &= 40\llb \big[V_{1} \Tijk23,4,5  - {1\over 11} s_{23}(A_{13|2,4,5} - A_{12|3,4,5})\big]\, , \cr
}$$
where
\eqn\Aijkdef{
A_{12|3,4,5} \equiv {1\over 6}[V^1 (A^2\g^{mn}W^3) - V^2 (A^1\g^{mn}W^3)](\l\g_m W^4)(\l\g_n W^5) + (3\leftrightarrow 4,5).
}
Note that the admixtures of \Aijkdef\ drop out from BRST invariant combinations of \TodEquiv\ such as $\langle {T_{1|23,4,5}\over s_{23}}+
{T_{12|3,4,5}\over s_{12}}-
{T_{13|2,4,5}\over s_{13}}
\rangle$, that is why the amplitudes obtained from the minimal \oneloopbb\ and the non-minimal pure
spinor formalism \oneloopNMPS\ agree.

\appendix{C}{Symmetries of two-loop kinematic factors}
\applab\appJac

\noindent This appendix collects the superspace manipulations responsible for some of the symmetry properties of
the two-loop scalar and vectorial kinematic factors.

\subsec{The Jacobi-like identity of scalar kinematic factors}

\noindent The kinematic factor \MDs\
$$\eqalignno{
T_{A,B|C,D}(\lambda,\lb) &={2\over\llb^6} (\lb\g_{m_1n_1p_1}r)(\lb\g_{def}r)(\lb\g_{m_2n_2p_2}r)(\l\g^{m_1defm_2}\l)\cr
 &\quad\times\big[ (\l\g^{n_1} W_{A})(\l\g^{p_1} W_B)(\l\g^{n_2} W_C)(\l\g^{p_2}W_D)\big]\,,\cr
}$$
is now demonstrated to satisfy the identity
\eqn\JacobiT{
T_{A,B|C,D}(\lambda,\lb) + T_{A,D|B,C}(\lambda,\lb) + T_{A,C|D,B}(\lambda,\lb) = 0.
}
To see this one uses the gamma matrix identity
\eqn\gatre{
(\lb\g_{def}r)(\l\g^{m_1defm_2}\l) = 48\llb (\l\g^{m_1}\g^{m_2}r) - 48(\l\g^{m_1}\g^{m_2}\lb)(\l r)
}
together with $(\lb\g_{m_2n_2p_2}r) = (\lb\g^{m_2}\g^{n_2}\g^{p_2}r)$ and $(\lb \g^{m_2})_\a (\lb\g_{m_2})_\b = 0$ to obtain
\eqnn\tmpJac
$$\eqalignno{
(\lb\g_{m_2n_2p_2}r)(\lb\g_{def}r)(\l\g^{m_1defm_2}\l)
 &=  48\llb (\lb\g_{m_2n_2p_2}r)(\l\g^{m_1}\g^{m_2}r)&\tmpJac\cr
 &= 48\llb(\lb\g^{m_2}\g^{m_1}\l)(r\g^{m_2n_2p_2}r)\, ,
}$$
where the cyclic identity $\g^{m_2}_{\a(\b}\g^{m_2}_{\g\d)} = 0$ and the constraint $(\lb\g^m r)= 0$
were used to arrive at the second line.
Therefore,
\eqn\fijac{
(\lb\g_{m_1n_1p_1}r)(\lb\g_{m_2n_2p_2}r)(\lb\g_{def}r)(\l\g^{m_1defm_2}\l) =96\llb^2(\lb\g^{an_1p_1}r)(r\g^{an_2p_2}r) \ .
}
After using \fijac, the identity \JacobiT\ follows by noting that it is equivalent to
\eqn\showeq{
(\lb\g^{an_1p_1}r)(r\g^{an_2p_2}r) + (\lb\g^{a p_2 p_1}r)(r\g^{a n_1 n_2}r) + (\lb\g^{an_2 p_1}r)(r\g^{ap_2 n_1}r) = 0 \ ,
}
and \showeq\ can be shown using $\g^{a}_{\a(\b}\g^{a}_{\g\d)} = 0$.

\subsec{Relating vector kinematic factors}

In order to prove the symmetry \Smfinal\ of the vectorial kinematic factor in \Sone\ and \Stwo, the pure
spinor constraint can be invoked to decompose the gamma matrices
in the factor 
$$(W_A\g^{a_1a_2a_3}W_B)(\l\g^{a_2m_1n_1p_1m_2}\l) = -(W_A\g^{a_2}\g^{a_1}\g^{a_3}W_B)
(\l\g^{a_2}\g^{m_1}\g^{n_1}\g^{p_1}\g^{m_2}\l)
$$
contained in \Sone. The identity
$$\displaylines{
(W_A\g^{a_2}\g^{a_1}\g^{a_3}W_B)(\l\g^{a_2}\g^{m_1}\g^{n_1}\g^{p_1}\g^{m_2}\l)
=-(W_A\g^{a_2}\g^{m_1}\g^{n_1}\g^{p_1}\g^{m_2}\l)(\l\g^{a_2}\g^{a_1}\g^{a_3}W_B)\cr
 -(W_A\g^{a_2}\l)(\l\g^{m_2}\g^{p_1}\g^{n_1}\g^{m_1} \g^{a_2}\g^{a_1}\g^{a_3}W_B)
}$$
then allows applications of the pure spinor constraint in the form of $(\l\g_r)_\a(\l\g^r)_\b=0$;
ultimately leading to
$$
S^{(1)\,m}_{A,B,C|D,E}(\lambda,\lb) = S^{(2)\,m}_{A,B,C|D,E}(\lambda,\lb) + S^{(2)\,m}_{A,D,E|B,C}(\lambda,\lb)
 + S^{(2)\,m}_{B,D,E|A,C}(\lambda,\lb) + S^{(2)\,m}_{C,D,E|A,B}(\lambda,\lb)
 $$
which implies \Smfinal.

\listrefs

\bye

%% file: harvmacM.tex

%
\def\unredoffs{}
\tolerance=1000\hfuzz=2pt
\catcode`\@=11 
\ifx\hyperdef\UNd@FiNeD\def\hyperdef#1#2#3#4{#4}\def\hyperref#1#2#3#4{#4}\def\href#1#2{#2}\fi
\magnification=1200\unredoffs\baselineskip=16pt plus 2pt minus 1pt
\def\Date#1{\vfill\leftline{#1}\tenpoint\supereject%
\footline={\hss\tenrm\hyperdef\hypernoname{page}\folio\folio\hss}}%

{\count255=\time\divide\count255 by 60 \xdef\hourmin{\number\count255}
 \multiply\count255 by-60\advance\count255 by\time
 \xdef\hourmin{\hourmin:\ifnum\count255<10 0\fi\the\count255}
}
\def\date{\number\day.\number\month.\number\year\ at \hourmin}


\def\nolabels{\def\wrlabeL##1{}\def\eqlabeL##1{}\def\reflabeL##1{}}
\def\writelabels{\def\wrlabeL##1{\leavevmode\vadjust{\rlap{\smash%
{\line{{\escapechar=` \hfill\rlap{\sevenrm\hskip.03in\string##1}}}}}}}%
\def\eqlabeL##1{{\escapechar-1\rlap{\sevenrm\hskip.05in\string##1}}}%
\def\reflabeL##1{\noexpand\llap{\noexpand\sevenrm\string\string\string##1}}}
\nolabels

\global\newcount\secno \global\secno=0
\global\newcount\meqno \global\meqno=1
\def\s@csym{}

\def\newsec#1\par{\global\advance\secno by1%
{\toks0{#1}\message{(\the\secno. \the\toks0)}}%
\global\subsecno=0\eqnres@t\let\s@csym\secsym\xdef\secn@m{\the\secno}\noindent
{\bf\hyperdef\hypernoname{section}{\the\secno}{\the\secno.} #1}%
\writetoca{{\string\hyperref{}{section}{\the\secno}{\bf \the\secno\quad}} {\bf #1}}\par%
\nobreak\medskip\nobreak\noindent\ignorespaces}
\def\eqnres@t{\xdef\secsym{\the\secno.}\global\meqno=1\bigbreak\bigskip}
\def\sequentialequations{\def\eqnres@t{\bigbreak}}\xdef\secsym{}

\global\newcount\subsecno \global\subsecno=0
\def\subsec#1\par{\global\advance\subsecno by1%
{\toks0{#1}\message{(\s@csym\the\subsecno. \the\toks0)}}%
\global\subsubsecno=0%
\ifnum\lastpenalty>9000\else\bigbreak\fi
\noindent{\it\hyperdef\hypernoname{subsection}{\secn@m.\the\subsecno}%
{\secn@m.\the\subsecno.} #1}\writetoca{\string\hskip1.45cm
{\string\hyperref{}{subsection}{\secn@m.\the\subsecno}{\secn@m.\the\subsecno.}}
{#1}}\par\nobreak\medskip\nobreak\noindent\ignorespaces}

\def\appendix#1#2{\global\meqno=1\global\subsecno=0\xdef\secsym{\hbox{#1.}}%
\bigbreak\bigskip\noindent{\bf Appendix \hyperdef\hypernoname{appendix}{#1}%
{#1.} #2}{\toks0{(#1. #2)}\message{\the\toks0}}%
\xdef\s@csym{#1.}\xdef\secn@m{#1}%
\writetoca{{\string\hyperref{}{appendix}{#1}{\bf {#1}\quad}} {\bf #2}}%
\par\nobreak\medskip\nobreak}

%
\def\checkm@de#1#2{\ifmmode{\def\f@rst##1{##1}\hyperdef\hypernoname{equation}%
{#1}{#2}}\else\hyperref{}{equation}{#1}{#2}\fi}
\def\eqnn#1{\DefWarn#1\xdef #1{(\noexpand\relax\noexpand\checkm@de%
{\s@csym\the\meqno}{\secsym\the\meqno})}%
\wrlabeL#1\writedef{#1\leftbracket#1}\global\advance\meqno by1}
\def\f@rst#1{\c@t#1a\em@ark}\def\c@t#1#2\em@ark{#1}
\def\eqna#1{\DefWarn#1\wrlabeL{#1$\{\}$}%
\xdef #1##1{(\noexpand\relax\noexpand\checkm@de%
{\s@csym\the\meqno\noexpand\f@rst{##1}1}{\hbox{$\secsym\the\meqno##1$}})}
\writedef{#1\numbersign1\leftbracket#1{\numbersign1}}\global\advance\meqno by1}
\def\eqn#1#2{\DefWarn#1%
\xdef #1{(\noexpand\hyperref{}{equation}{\s@csym\the\meqno}%
{\secsym\the\meqno})}$$#2\eqno(\hyperdef\hypernoname{equation}%
{\s@csym\the\meqno}{\secsym\the\meqno})\eqlabeL#1$$%
\writedef{#1\leftbracket#1}\global\advance\meqno by1}
\def\xeqn{\expandafter\xe@n}\def\xe@n(#1){#1}
\def\xeqna#1{\expandafter\xe@n#1}
\def\eqns#1{(\e@ns #1{\hbox{}})}
\def\e@ns#1{\ifx\UNd@FiNeD#1\message{eqnlabel \string#1 is undefined.}%
\xdef#1{(?.?)}\fi{\let\hyperref=\relax\xdef\next{#1}}%
\ifx\next\em@rk\def\next{}\else%
\ifx\next#1\xeqn#1\else\def\n@xt{#1}\ifx\n@xt\next#1\else\xeqna#1\fi
\fi\let\next=\e@ns\fi\next}

\def\DefWarn#1{\ifx\UNd@FiNeD#1\else
\immediate\write16{*** WARNING: the label \string#1 is already defined ***}\fi}
%
\newskip\footskip\footskip14pt plus 1pt minus 1pt 
\def\footnotefont{\ninepoint}\def\f@t#1{\footnotefont #1\@foot}
\def\f@@t{\baselineskip\footskip\bgroup\footnotefont\aftergroup\@foot\let\next}
\setbox\strutbox=\hbox{\vrule height9.5pt depth4.5pt width0pt}
\global\newcount\ftno \global\ftno=0
\def\foot{\global\advance\ftno by1\def\foot@rg{\hyperref{}{footnote}%
{\the\ftno}{\the\ftno}\xdef\foot@rg{\noexpand\hyperdef\noexpand\hypernoname%
{footnote}{\the\ftno}{\the\ftno}}}\footnote{$^{\foot@rg}$}}
%
%
%
\global\newcount\refno \global\refno=1
\newwrite\rfile
\def\ref{[\hyperref{}{reference}{\the\refno}{\the\refno}]\nref}
\def\nref#1{\DefWarn#1%
\xdef#1{[\noexpand\hyperref{}{reference}{\the\refno}{\the\refno}]}%
\writedef{#1\leftbracket#1}%
\ifnum\refno=1\immediate\openout\rfile=\jobname.refs\fi
\chardef\wfile=\rfile\immediate\write\rfile{\noexpand\item{[\noexpand\hyperdef%
\noexpand\hypernoname{reference}{\the\refno}{\the\refno}]\ }%
\reflabeL{#1\hskip.31in}\pctsign}\global\advance\refno by1\findarg}
\def\findarg#1#{\begingroup\obeylines\newlinechar=`\^^M\pass@rg}
{\obeylines\gdef\pass@rg#1{\writ@line\relax #1^^M\hbox{}^^M}%
\gdef\writ@line#1^^M{\expandafter\toks0\expandafter{\striprel@x #1}%
\edef\next{\the\toks0}\ifx\next\em@rk\let\next=\endgroup\else\ifx\next\empty%
\else\immediate\write\wfile{\the\toks0}\fi\let\next=\writ@line\fi\next\relax}}
\def\striprel@x#1{} \def\em@rk{\hbox{}}
\def\lref{\begingroup\obeylines\lr@f}
\def\lr@f#1#2{\DefWarn#1\gdef#1{\let#1=\UNd@FiNeD\ref#1{#2}}\endgroup\unskip}
\def\semi{;\hfil\break}
\def\addref#1{\immediate\write\rfile{\noexpand\item{}#1}} 
\def\listrefs{\vfill\supereject\immediate\closeout\rfile\writestoppt
\baselineskip=\footskip\centerline{{\bf References}}\bigskip{\parindent=20pt%
\frenchspacing\escapechar=` \input \jobname.refs\vfill\eject}\nonfrenchspacing}
\def\startrefs#1{\immediate\openout\rfile=\jobname.refs\refno=#1}
\def\xref{\expandafter\xr@f}\def\xr@f[#1]{#1}
\def\refs#1{\count255=1[\r@fs #1{\hbox{}}]}
\def\r@fs#1{\ifx\UNd@FiNeD#1\message{reflabel \string#1 is undefined.}%
\nref#1{need to supply reference \string#1.}\fi%
\vphantom{\hphantom{#1}}{\let\hyperref=\relax\xdef\next{#1}}%
\ifx\next\em@rk\def\next{}%
\else\ifx\next#1\ifodd\count255\relax\xref#1\count255=0\fi%
\else#1\count255=1\fi\let\next=\r@fs\fi\next}
%

%
\newwrite\ffile\global\newcount\figno \global\figno=1
\def\fig{fig.~\hyperref{}{figure}{\the\figno}{\the\figno}\nfig}
\def\nfig#1{\DefWarn#1%
\xdef#1{fig.~\noexpand\hyperref{}{figure}{\the\figno}{\the\figno}}%
\writedef{#1\leftbracket fig.\noexpand~\xfig#1}%
\ifnum\figno=1\immediate\openout\ffile=\jobname.figs\fi\chardef\wfile=\ffile%
{\let\hyperref=\relax
\immediate\write\ffile{\noexpand\medskip\noexpand\item{Fig.\ %
\noexpand\hyperdef\noexpand\hypernoname{figure}{\the\figno}{\the\figno}. }
\reflabeL{#1\hskip.55in}\pctsign}}\global\advance\figno by1\findarg}
\def\xfig{\expandafter\xf@g}\def\xf@g fig.\penalty\@M\ {}
\def\figs#1{figs.~\f@gs #1{\hbox{}}}
\def\f@gs#1{{\let\hyperref=\relax\xdef\next{#1}}\ifx\next\em@rk\def\next{}\else
\ifx\next#1\xfig #1\else#1\fi\let\next=\f@gs\fi\next}
%
\def\figin{\epsfcheck\figin}\def\figins{\epsfcheck\figins}
\def\epsfcheck{\ifx\epsfbox\UnDeFiNeD
\message{(NO epsf.tex, FIGURES WILL BE IGNORED)}
\gdef\figin##1{\vskip2in}\gdef\figins##1{\hskip.5in}
\else\message{(FIGURES WILL BE INCLUDED)}%
\gdef\figin##1{##1}\gdef\figins##1{##1}\fi}
\def\DefWarn#1{}
\def\figinsert{\goodbreak\topinsert}
\def\ifig#1#2#3{\DefWarn#1\xdef#1{fig.~\the\figno}
\writedef{#1\leftbracket fig.\noexpand~\the\figno}%
\figinsert\figin{\centerline{#3}}
\smallskip
\leftskip=20pt \rightskip=20pt
\baselineskip12pt\noindent
{{\bf Fig.~\the\figno}\ \ninepoint #2}
\medskip
\global\advance\figno by1\par\endinsert}
\newwrite\lfile
{\escapechar-1\xdef\pctsign{\string\%}\xdef\leftbracket{\string\{}
\xdef\rightbracket{\string\}}\xdef\numbersign{\string\#}}
\def\writedefs{\immediate\openout\lfile=label.defs \def\writedef##1{%
{\let\hyperref=\relax\let\hyperdef=\relax\let\hypernoname=\relax
 \immediate\write\lfile{\string\def\string##1\rightbracket}}}}%
\def\writestop{\def\writestoppt{\immediate\write\lfile{\string\pageno
 \the\pageno\string\startrefs\leftbracket\the\refno\rightbracket
 \string\def\string\secsym\leftbracket\secsym\rightbracket
 \string\secno\the\secno\string\meqno\the\meqno}\immediate\closeout\lfile}}
\def\writestoppt{}\def\writedef#1{}

\def\seclab#1{\DefWarn#1%
\xdef #1{\noexpand\hyperref{}{section}{\the\secno}{\the\secno}}%
\writedef{#1\leftbracket#1}\wrlabeL{#1=#1}}
\def\subseclab#1{\DefWarn#1%
\xdef #1{\noexpand\hyperref{}{subsection}{\the\secno.\the\subsecno}%
{\the\secno.\the\subsecno}}\writedef{#1\leftbracket#1}\wrlabeL{#1=#1}}
\def\applab#1{\DefWarn#1%
\xdef #1{\noexpand\hyperref{}{appendix}{\secn@m}{\secn@m}}%
\writedef{#1\leftbracket#1}\wrlabeL{#1=#1}}
\newwrite\tfile \def\writetoca#1{}
\def\leaderfill{\leaders\hbox to 1em{\hss.\hss}\hfill}
\def\writetoc{\immediate\openout\tfile=\jobname.toc
   \def\writetoca##1{{\edef\next{\write\tfile{\noindent ##1
   \string\leaderfill{
   \string\hyperref{}{page}{\noexpand\number\pageno}%
   {\noexpand\number\pageno}} \par}}\next}}
}
\newread\ch@ckfile
\def\listtoc{\immediate\closeout\tfile\immediate\openin\ch@ckfile=\jobname.toc
\ifeof\ch@ckfile\message{no file \jobname.toc, no table of contents this pass}%
\else\closein\ch@ckfile\centerline{\bf Contents}\nobreak\medskip%
{\baselineskip=16pt\footnotefont\parskip=0pt\catcode`\@=11\input\jobname.toc
\catcode`\@=12\bigbreak\bigskip}\fi}
\catcode`\@=12 
\def\tenpoint{\def\rm{\fam0\tenrm}
\textfont0=\tenrm \scriptfont0=\sevenrm \scriptscriptfont0=\fiverm
\textfont1=\teni  \scriptfont1=\seveni  \scriptscriptfont1=\fivei
\textfont2=\tensy \scriptfont2=\sevensy \scriptscriptfont2=\fivesy
\textfont\itfam=\tenit \def\it{\fam\itfam\tenit}\def\footnotefont{\ninepoint}%
\textfont\bffam=\tenbf \def\bf{\fam\bffam\tenbf}\def\sl{\fam\slfam\tensl}\rm}
\font\ninerm=cmr9 \font\sixrm=cmr6 \font\ninei=cmmi9 \font\sixi=cmmi6
\font\ninesy=cmsy9 \font\sixsy=cmsy6 \font\ninebf=cmbx9
\font\nineit=cmti9 \font\ninesl=cmsl9 \skewchar\ninei='177
\skewchar\sixi='177 \skewchar\ninesy='60 \skewchar\sixsy='60
\def\ninepoint{\def\rm{\fam0\ninerm}
\textfont0=\ninerm \scriptfont0=\sixrm \scriptscriptfont0=\fiverm
\textfont1=\ninei \scriptfont1=\sixi \scriptscriptfont1=\fivei
\textfont2=\ninesy \scriptfont2=\sixsy \scriptscriptfont2=\fivesy
\textfont\itfam=\ninei \def\it{\fam\itfam\nineit}\def\sl{\fam\slfam\ninesl}%
\textfont\bffam=\ninebf \def\bf{\fam\bffam\ninebf}\rm}
%
\hyphenation{anom-aly anom-alies coun-ter-term coun-ter-terms}

\global\newcount\subsubsecno \global\subsubsecno=0
\def\subsubsec#1\par{\global\advance\subsubsecno by1%
{\toks0{#1}\message{(\the\secno\the\subsecno\the\subsubsecno. \the\toks0)}}%
\ifnum\lastpenalty>9000\else\bigbreak\fi
\noindent{\it\hyperdef\hypernoname{subsubsection}{\the\secno.\the\subsecno\the\subsubsecno}%
{\the\secno.\the\subsecno.\the\subsubsecno.} #1}
\par\nobreak\medskip\nobreak\noindent\ignorespaces}

\def\DefWarn#1{}
\def\tikzcaption#1#2{\DefWarn#1\xdef#1{Fig.~\the\figno}
\writedef{#1\leftbracket Fig.\noexpand~\the\figno}%
{
\smallskip
\leftskip=20pt \rightskip=20pt \baselineskip12pt\noindent
{{\bf Fig.~\the\figno}\ \ninepoint #2}
\bigskip
\global\advance\figno by1 \par}}

\def\ntoalpha#1{%
\ifcase#1%
@%
\or A\or B\or C\or D\or E\or F\or G\or H\or I
\fi
}

\global\newcount\appno \global\appno=1
\def\applab#1{\xdef #1{\ntoalpha\appno}\writedef{#1\leftbracket#1}\wrlabeL{#1=#1}
\global\advance\appno by1}

\def\preprint#1 #2\par{\rightline{\vbox{\baselineskip12pt\hbox{#1}\hbox{#2}}}\vskip2cm}
%
\def\title#1\par{\centerline{\bf #1}\nopagenumbers\pageno=0}
\def\author#1\par{\bigskip\bigskip\centerline{#1}}

\newcount\addressno

\def\email#1#2{\unskip$^#1$\footnote{\null}{\kern-\parindent \llap{$^#1$\hskip1pt}email: #2}}

\def\startcenter{%
  \par
  \begingroup
  \leftskip=0pt plus 1fil
  \rightskip=\leftskip
  \parindent=0pt
  \parfillskip=0pt
}
\def\stopcenter{\endgroup}

\def\address{\bigskip%
  \ifnum\the\addressno=0\else\stopcenter\endgroup\fi
  \advance\addressno by 1%
  \begingroup
  \startcenter
  \it
  \obeylines
  \addressAux
}
\def\addressAux#1{#1}

\def\abstract{\stopcenter\endgroup\bigskip\bigskip\noindent}

\def\Dsl{\,\raise.15ex\hbox{/}\mkern-13.5mu D} 
\def\dsl{\raise.15ex\hbox{/}\kern-.57em\partial}
 
\def\boxeqn#1{\vcenter{\vbox{\hrule\hbox{\vrule\kern3pt\vbox{\kern3pt
	\hbox{${\displaystyle #1}$}\kern3pt}\kern3pt\vrule}\hrule}}}


\def\ap{{\alpha^{\prime}}}
\def\halfap#1{\Big({\ap\over 2}\Big)^{\mkern-4mu #1}}
\def\a{\alpha}
\def\b{{\beta}}
\def\g{{\gamma}}
\def\d{{\delta}}
\def\e{{\epsilon}}
\def\l{\lambda}
\def\k{{\kappa}}
\def\s{{\sigma}}
\def\t{{\theta}}
\def\om{{\omega}}
\def\lb{{\overline\lambda}}
\def\llb{(\l\lb)}
\def\wb{{\overline w}}
\def\half{{1\over 2}}
\def\p{{\partial}}
\def\pb{{\overline\partial}}

\def\bar{\overline}
\def\({\left(}
\def\){\right)}

\def\ImOmega{\mathop{\rm Im}\Omega}

\def\Im{\mathop{{\rm Im}}} 


\def\qed{\hbox{\hskip 3pt
\vbox{\hrule\hbox to 7pt{\vrule height 7pt\hfill\vrule}
\hrule}}\hskip3pt}

\overfullrule=0pt\relax

\frenchspacing

\newread\instream \openin\instream= label.defs
\ifeof\instream \message{No labels in advance yet. Wait till next pass.}
\else \closein\instream \input label.defs
\fi
\writedefs

\def\arXiv:#1].{\hepthStrip#1 \nil}
\def\hepthStrip#1 #2\nil{\href{http://arxiv.org/abs/#1}{arXiv:#1 #2\unskip}].}